\documentclass[lettersize,journal]{IEEEtran}
\usepackage{cite,amssymb}
\usepackage[pdftex]{graphicx}
\usepackage[cmex10]{amsmath}
\usepackage{algorithmicx}   
\usepackage{algorithm}
\usepackage{algpseudocode}
\usepackage{array,balance} 
\usepackage{stfloats}
\usepackage{caption}
\ifCLASSOPTIONcompsoc
\usepackage[caption=false,font=normalsize,labelfont=sf,textfont=sf]{subfig}
\else
\usepackage[caption=false,font=footnotesize]{subfig}
\fi
\usepackage{color, soul} 
\soulregister\cite7 
\soulregister\citep7 
\soulregister\citet7 
\soulregister\ref7 
\soulregister\pageref7 

\usepackage{float}
\usepackage{makecell} 
\usepackage{enumerate}
\usepackage{url}
\usepackage{multirow}
\usepackage{longtable}
\usepackage{graphicx}
\usepackage{amsfonts}
\usepackage{exscale}
\usepackage{relsize}
\usepackage{multicol,color,bm}

\usepackage{arydshln}

\makeatletter
\renewcommand{\maketag@@@}[1]{\hbox{\m@th\normalsize\normalfont#1}}%
\makeatother

\begin{document}

\title{Channel Charting based Fast Beam Tracking Design and Implementation}

\author{Jiawei Zhang, Shihan Wang, Jienan Chen\textsuperscript{*},~\IEEEmembership{Senior Member,~IEEE}, Fan Wu, Jiyun Tao, Zheqi Gu
\thanks{
Jiawei Zhang, Shihan Wang and Jienan Chen are with University of Electronic Science and Technology of China, Chengdu, 611731, China (e-mail: zhanggavei0426@gmail.com; wangshihan127@outlook.com; jesson.chen@outlook.com). \textsuperscript{*}Jienan Chen is the corresponding author.

Fan Wu is with Nanyang Technological University, Singapore, 639798, Singapore (e-mail: fan009@e.ntu.edu.sg)

Jiyun Tao and Zheqi Gu are with Huawei Technologies Co., Ltd, Chengdu, 610041, China (e-mail: jytaonone@hotmail.com; guzheqi@gmail.com)
}
        }



\maketitle

\begin{abstract}
In the beyond fifth-generation (B5G) and upcoming sixth-generation (6G) wireless communication systems, millimeter (mmWave) wave technology is a promising solution for offering additional bandwidth resources and mitigating spectrum congestion. Beam tracking is an essential procedure for providing reliable communication services in the mmWave communication system, with the challenge of providing consistent and accurate tracking performance. In this study, we introduce a low-overhead beam tracking algorithm based on channel charting, which significantly reduces beam scanning times during the tracking process. By projecting the beam information to the channel chart, the beam tracking problem is transformed into the acquisition of the beam cluster in the channel chart. Leveraging contrastive learning, the proposed channel chart projects high-dimensional channel state information into a low-dimensional feature space that preserves spatial proximities. Using a dynamic candidate beam acquisition strategy, the complexity of our beam tracking algorithm is significantly reduced. The proposed algorithm significantly reduces scanning complexity while maintaining high prediction accuracy, achieving an accuracy of 98.27\% in simulation environments. Compared to existing methods, the proposed method can reduce beam scanning times by up to 55.9\%. In addition, we also performed field tests, and the measured results demonstrated excellent communication quality during mobility.
\end{abstract}


\section{Introduction}
\IEEEPARstart{M}{illimeter-wave} (mmWave) communication is an emerging and promising technology in modern communication systems, driven by state-of-the-art techniques and ever-increasing demands. The abundant spectral resources at millimeter wave frequencies have the potential to enable communication data rates on the order of gigabits per second \cite{fast_beam_alignment}. Beamforming addresses the issue of severe propagation loss in mmWave communications by providing antenna gain and ensuring reliable connectivity \cite{noncooperative}.

Ensuring high efficiency in mmWave communication relies on precise beam alignment between transmitters and receivers. The 3rd Generation Partnership Project (3GPP) has standardized beam management protocols. To acquire channel state information (CSI), an initial search is required to identify the optimal beam pair. However, this phase is followed by a subsequent monitoring phase, which plays a more critical role in maintaining robust and stable communication. The frequent mobility of user equipment (UE) and dynamic environmental changes further underscore the necessity for continuous and adaptive beam monitoring. Beam tracking fundamentally operates as a repetitive beam alignment process to address time-varying channel conditions\cite{8458146}.  After the initial beam alignment, the base station (BS) sends pilot symbols using a few probe beams. BS and UE can use angles of arrival (AoA) and angles of departure (AoD) information acquired from prior beam searches to perform targeted beam scanning, significantly reducing the required beam configurations\cite{8757185}.

The accuracy of beam tracking problems has been investigated in some studies. The study in reference \cite{7063467} introduces a model for the evolution of channel parameters, utilizing a first-order Gaussian Markov process specifically tailored for the context of tracking. Further advancements are noted in reference \cite{7510902}, where an Extended Kalman Filter (EKF) is applied to achieve millimeter-wave beam tracking. This approach has shown higher accuracy compared to other Kalman-filter (KF)-based methods. Nevertheless, it also has drawbacks, such as the accumulation of errors and considerable overhead in tracking and sweeping.

To further reduce the complexity of beam training, research \cite{zhang2019codebook} proposes an efficient and low-complexity beam training scheme for millimeter-wave communication by multiple users. It adopts a greedy search approach to gradually narrow down the set of beam candidates, thereby improving the training efficiency. In scenarios involving large-scale antenna arrays, this method significantly reduces computational resource consumption. Research \cite{zhang2021training} cleverly reformulates the beam selection problem into a second-order cone programming (SOCP) and nonlinear optimization problem. The proposed optimization strategy significantly improves the accuracy of beam tracking algorithms.

The study in \cite{ning2021unified} proposes a unified 3D beam training and tracking scheme, designing a low-complexity 3D hierarchical codebook and training protocol, optimizing the beam alignment process. The newly designed hierarchical codebook and protocol strike a good balance between complexity and performance, and can dynamically adapt to different communication requirements.


The integration of machine learning techniques has further expanded the potential of beam tracking methods. Research \cite{liang2023millimetre} proposes a beam tracking method based on the fusion of machine learning and KF, with the objective of improving the accuracy and robustness of beam tracking in dynamic environments. It introduces variational Bayesian inference to convert the acquisition of the KF state transition matrix into the training of a neural network and uses long-short-term memory (LSTM) networks for beam angle prediction. This approach combines data-driven and model-driven methods, overcoming the traditional KF's dependence on linear assumptions. Research \cite {burghal2019machine,dehkordi2021adaptive} employs machine learning (ML) techniques to address beam tracking challenges. 


Some research has suggested that the integration of environmental and location data might offer viable solutions to major issues within mmWave systems, as detailed in \cite{o49}. Research\cite{o50} considers mapping from the position of users to the gain of the channel to guide beam tracking, which can greatly improve beamforming performance, particularly in scenarios that require switching due to obstructions. However, the above algorithm requires precise location information, which involves user privacy.

Channel charting facilitates the extraction of channel features associated with large-scale fading characteristics in wireless communication systems while maintaining the spatial consistency inherent to the channel \cite{ferrand_triplet-based_2021}\cite{9306087}. The studies in \cite{kazemi_channel_2022} and \cite{ponnada2021best} propose an algorithm for beam tracking assisted by channel charting. The algorithms utilize the channel covariance matrix to construct a channel chart and map the signal-to-noise ratio (SNR) of the optimal beam onto it. In results, the communication system can track the optimal beam between the UE and the BS with minimal beam scanning, achieving higher accuracy and lower scanning overhead. However, in practical communication systems, obtaining the channel covariance matrix requires channel estimation, which involves frequent signal exchanges between the BS and the UE, resulting in high overhead and latency. 





In this paper, we propose a low-complexity beam tracking framework that leverages unsupervised channel charting to achieve robust communication performance with minimized resource expenditure. By transforming high-dimensional CSI, including multi-path AoA and AoD features into a topology-preserving low-dimensional chart, our method establishes a geometry-aware latent representation of UE spatial relationships. This charting process enables dynamic beam cluster identification through an adaptive mapping table mechanism, where candidate beams are intelligently selected based on UE trajectory patterns rather than exhaustive beamspace scanning. The contributions of this work are summarized as follows:

\begin{enumerate}
    \item \textbf{Channel charting based dimension compression beam tracking algorithm framework}: We propose a data-driven beam tracking framework based on channel charting to reduce scanning overhead while maintaining precise beam prediction. The algorithm employs a triplet network to generate a channel chart, which compresses high-dimensional CSI into a lower-dimensional space. This approach not only captures essential spatial features but also preserves the original relative positional relationships. The trained channel charting smooths out noise and high-frequency variations, enabling the system to adapt more effectively to dynamic changes in mobile devices and the channel environment. By using the channel chart, a dynamically adjustable candidate beam set is generated, allowing for intelligent beam tracking with minimal computational cost and high accuracy.
    \item \textbf{Improved dynamic candidate beam set acquisition using mapping table theory}: We propose a method that can dynamically obtain the candidate beam set based on the theory of mapping tables. This strategy establishes a direct relationship between the channel chart and the set of candidate beams, allowing the real-time beam tracking algorithm to obtain candidate beams for antenna scanning. We have proven that the time complexity of this method is \( \mathcal{O}(1) \). This allows the algorithm to rapidly adapt to changes in the channel environment, ensuring high tracking accuracy while minimizing computational load. Compared to existing research, this method reduces neighborhood search times by half and reduces overall scanning times by 55.9\%.
    \item \textbf{Field-testing and implementation of the prototype}: We validate the performance of the proposed beam tracking algorithm through extensive simulations and field tests.  The experimental platform employs two 26-GHz mmWave antennas and a software-defined radio (SDR), emulating practical communication environments. Results demonstrate that the algorithm performs well in dynamic and NLOS scenarios, effectively capturing channel angle variations and achieving low-complexity and high-robustness real-time beam tracking.
\end{enumerate}

A preliminary version of this work was previously published in \cite{our-pre-work}.
In comparison, this work has enhanced the precision and improved the algorithm architecture, allowing the generated channel chart to better maintain relative positional relationships. Additionally, the complexity of retrieving candidate beam sets has been reduced by utilizing a mapping table, cutting the number of neighborhood searches in half, thereby making the algorithm more efficient and robust. We have also discussed the performance of the algorithm under different parameters and conducted a simulation analysis of the algorithm's timeliness under 3GPP scenarios.

The rest of this article is organized as follows. Section II presents the system model and the problem formulation at a high level. Section III details the processes of the proposed beam tracking algorithm, while Section IV provides an overview of how the algorithm works and describes the prediction error. Numerical simulations are presented in Section V, and conclusions are drawn in Section VI.

Notations: Bold uppercase $\mathbf{A}$ and bold lowercase $\mathbf{a}$ denote
 matrices and column vectors, respectively. Without particular
 specification, non-bold letters $A,a$ denote scalars. $\mathbf{b}$ denotes the beam pair, while $b$ denotes the beam index in this paper. $\mathbb{E}(\cdot)$ and $(\cdot)^{H}$ denote the mathematical
 expectation and Hermitian operators, respectively. In addition, $\left|\cdot\right|$ and $\left\| \cdot\right\|$ denote the magnitude of a complex number and $\ell_2$-norm of a vector. Moreover, $\mathcal{O}$ and $\operatorname{mod}(\cdot)$ denote the order of complexity and the operation of modulo, respectively. 

\begin{table*}[t]
\centering

\caption{List of Main Notations}
\scriptsize
\renewcommand{\arraystretch}{0.85}
\label{tab:notations}
\begin{tabular}{|l|p{6cm}|l|p{6cm}|}
\hline
\textbf{Notation} & \textbf{Description} & \textbf{Notation} & \textbf{Description} \\ \hline
$N_t$ & BS antenna count & $N_r$ & UE antennas count \\ \hline
$t$ & Time slot index & $l$ & Path index \\ \hline
$\mathbf{H}(t)$& Time-varying channel matrix& $\beta_{l}(t)$& Complex gain of the $l$-th path\\ \hline
$\phi^{(t)}_l$, $\theta^{(t)}_l$& Spatial angles of departure, $\theta^{(t)}_l = \sin(\phi^{(t)}_l)$& $\phi^{(r)}_l$,$\theta^{(r)}_l$& Spatial angles of arrival, $\theta^{(r)}_l = \sin(\phi^{(r)}_l)$\\ \hline
$\mathbf{a^{(t)}}$& BS array response vector & $\mathbf{a^{(r)}}$& UE array response vector \\ \hline

$\mathbf{C}$& Beamforming vectors&$\mathbf{f}_{n}$,$\mathbf{w}_{m}$& precoder and combiner \\ \hline

$a_{mn}(\alpha, \phi)$& The response of the element at position $(x_m, y_n)$ & $\mathbf{a_p}(\alpha,\phi)$ & Collection of the responses of all antenna elements\\ \hline
$y_{m, n}$& The received signal& $x$& Pilot symbol \\ \hline

$h_{m,n}(t)$& The
effective channel coefficient& $\tilde{m},\tilde{n}$& The best beam pair index \\ \hline

$\Gamma_{m,l}^{(r)} $& The beam alignment metric &  $g_{\theta^{}}^{e c}(\cdot),g_{\theta^{}}^{d c}(\cdot)$ & Encoder/decoder network \\ \hline
$\Phi_{cc}$& Channel charting based beam prediction strategy
& $t_e$& The time length\\ \hline
$t_d$& The delay of the historical CSI& $\tau_t$& The timestamp information at time $t$\\ \hline

$\mathbf{x}_t$& Input information& $\mathbf{y}_t$& 2D channel chart feature corresponding to $\mathbf{x}_t$ \\ \hline

$\mathcal{H}$& Mapping function& $\mathcal{B}_{t} $& The predicted  scanning beam set\\ \hline
$\mathcal{W}_{t+1}$& The currently best beam pair indexes& $\mathbf{\bar{X}}$& Raw Second-order Moment\\ \hline
$\Omega$ & The path loss factor& $\sigma$& Path
loss paramete\\ \hline

$M', D'$& The dimensionality of CSI and the channel
chart & $\mathcal{C}_{n}$&  The neighborhood loss error\\ \hline

$\mathcal{T}$& The total number of training samples & $\mathcal{C}_{r}$&  The network’s decoding loss\\ \hline
$U$& Set of all possible keys & $m$&  Number of slots in the mapping table\\ \hline
$\left|U\right|$& Size of the set of keys & $n$&  Total number of slots in the mapping table\\ \hline
$ky$& Key generated for a coordinate & $K$&  Key generation function \\ \hline
$c$& Generating factor for the key generation& $T_{j}$ &  Linked list in the mapping table \\ \hline
$\mathcal{B}_t$ & Candidate beam set corresponding to $\mathbf{x}_t$ & $E_{ps}$&  The algorithm positioning error\\ \hline
$E_{pd}$& The total prediction
error & $N_{s}$&  The total number
of searches \\ \hline
$\zeta^{(t)},\zeta^{(r)}$& The number of neighborhood searches at both the transmitter and the receiver & $\mathcal{B}_{t,n},\mathcal{B}_{r,n}$&  The sizes of the candidate beam sets at time n for the transmitter or receiver
of searches \\ \hline
\end{tabular}

\end{table*}

\section{System Model and Background}
In this section, we describe the beam tracking system model and the basis of channel charting.
\subsection{System Model}
Consider a single-user downlink mmWave system in which a base station equipped with $N_t$ antennas communicates with user equipment having $N_r$ antennas. The time-varying channel matrix $\mathbf{H}(t) \in \mathbb{C}^{N_r \times N_t}$ follows the sparse multipath angular domain model \cite{lim2021deep} \cite{brady2013beamspace}:
\begin{equation}
\mathbf{H}(t) = \sum_{l=1}^{L} \beta_{l}(t) \mathbf{a^{(r)}}(\theta^{(r)}_l) \mathbf{a^{(t)}}(\theta^{(t)}_l)^H,
\end{equation}
where $L$ denotes the number of dominant propagation paths, $\beta_{l}(t)$ represents the complex gain of the $l$-th path, and $\theta^{(t)}_l = \sin(\phi^{(t)}_l)$, $\theta^{(r)}_l = \sin(\phi^{(r)}_l)$, $\phi^{(t)}_l$ and $\phi^{(r)}_l$ respectively characterize the spatial AoD and AoA. Without loss of generality, we assume the antenna spacing to be half of a wavelength \cite{karacora2023event}. Hence, the array response vectors are defined as:
\begin{equation}
\mathbf{a\left(\theta\right)} = \frac{1}{\sqrt{N}} \left[1, e^{j2\pi d \frac{\theta}{\lambda}}, ..., e^{j2\pi(N-1)d \frac{\theta}{\lambda}}\right]^T.
\end{equation}
where $N$ denotes the array length. Aiming at beam tracking, the BS and UE employ DFT-based codebooks for beam alignment. Let $N_t$ and $N_r$ denote the sizes of the candidate beam sets at the BS and UE, respectively. The beamforming vectors $\mathbf{C}=\left[\mathbf{c}_{0}, \mathbf{c}_{1}, \ldots, \mathbf{c}_{N-1}\right]$ are generated as:
\begin{equation}
\mathbf{c}_{k}=\frac{1}{\sqrt{N}}\left[1, e^{j 2 \pi \frac{k}{N}}, e^{j 2 \pi \frac{2 k}{N}}, \ldots, e^{j 2 \pi \frac{(N-1) k}{N}}\right]^{T},
\end{equation}
where $k\in[ 0, 1,\cdots, N-1]$, $N=N_t$ for the BS codebook $\left\{\mathbf{f}_{n}\right\}_{n=1}^{N_{t}}$, $\mathbf{f}_{n} \in \mathbb{C}^{N_{t} \times 1}$, and $N=N_r$ for the UE codebook $\left\{\mathbf{w}_{m}\right\}_{m=1}^{N_{r}}$ , $\mathbf{w}_{m} \in \mathbb{C}^{N_{r} \times 1}$.

Without loss of generality, the model can be extended to the case of a planar array, where the angles encompass both the azimuth and elevation angles. For a plane wave signal, the response factor of a single antenna element can be expressed as:
\begin{equation}
    a_{mn}(\alpha,\phi)=e^{jk(x_m\sin\phi\cos\alpha+y_n\sin\phi\sin\alpha)},
\end{equation}
where $k=\frac{2\pi}\lambda $, $\lambda$ is the wavelength of the signal. $(x_m, y_n)
$ are the coordinates of the antenna elements in the planar array.
The response vector of the entire planar array $\mathbf{a_p}(\alpha,\phi)$ is the collection of the responses of all the antenna elements:
\begin{equation}
    \mathbf{a_p}(\alpha, \phi) = \left[ 
        a_{mn}(\alpha, \phi) \mid 
        \begin{aligned}
            &m \in \{0,\dots,M-1\}, \\
            &n \in \{0,\dots,N-1\}
        \end{aligned}
    \right]^T.
\end{equation}

A codebook index represents a pair of beams. The model proposed in this paper is applicable to predict a single angle or a pair of angles.

During beam tracking, the BS transmits pilot symbol $x$ using $\mathbf{f}_n$ while the UE employs the combining vector $\mathbf{w}_m$. The received signal is:
 \begin{equation}
 y_{m, n}=\mathbf{w}_{m}^{H} \mathbf{H}(t) \mathbf{f}_{n} x+\mathbf{w}_{m}^{H} \mathbf{n}={h}_{m, n} x+\mathbf{w}_{m}^{H}\mathbf{n},
\end{equation}
where $\mathbf{f}_n$ and $\mathbf{w}_{m}$ are the precoder and combiner, respectively, and $\mathbf{n}$ is the Gaussian white noise during transmission. The effective channel coefficient $h_{m,n}(t)$ explicitly expands to:
\begin{equation}
h_{m,n}(t) = \sum_{l=1}^L \beta_l(t) \underbrace{\mathbf{w}_m^H \mathbf{a}^{(r)}(\theta^{(r)}_l)}_{\Gamma_{m,l}^{(r)}} \underbrace{\mathbf{a}^{(t)}(\theta^{(t)}_l)^H \mathbf{f}_n}_{\Gamma_{n,l}^{(t)}}.
\end{equation}

The beam alignment metric $\Gamma_{m,l}^{(r)} = \mathbf{w}_m^H\mathbf{a}^{(r)}(\theta^{(r)}_l) = \mathbf{a}^{(r)}(\theta^{(r)}_m)^H \mathbf{a}^{(r)}(\theta^{(r)}_l)$ achieves maximum gain when $\theta^{(r)}_m = \theta^{(r)}_l$, exploiting the orthogonality of DFT steering vectors.

To obtain the best beam pair, the BS needs to scan the beam space. The received signal matrix is:
\begin{equation}
\mathbf{Y}=\mathbf{W}^{H} \mathbf{H}(t) \mathbf{F}+\mathbf{W}^{H} \mathbf{N},
\end{equation}
where $\mathbf{W}=\left[\mathbf{w}_{\tilde{m}},\cdots,\mathbf{w}_{\tilde{M}}\right]$, 
$\mathbf{w}_{\tilde{m}} \in\left\{\mathbf{w}_{m}\right\}_{m=1}^{N_{r}}$, $\mathbf{F}=\left[\mathbf{f}_{\tilde{n}}, \cdots, \mathbf{f}_{\tilde{N}}\right]$,
$\mathbf{f}_{\tilde{n}} \in\left\{\mathbf{f}_{n}\right\}_{n=1}^{N_{t}}$ represent the beam angle set currently scanned by the BS and the UE. Note that when the scan size satisfies $\tilde{M}=N_{r}$, $\tilde{N}=N_{t}$, the sweeping process is equal to exhaustive sweeping. 

\subsection{Problem Formulation}
Based on the system model described above, we can formalize the beam tracking problem.
At each time instant, the base station employs a precoder $\mathbf{f}_{n} \in \mathbb{C}^{N_t \times 1}$ selected from the DFT-based codebook $\left\{\mathbf{f}_{n}\right\}_{n=1}^{N_{t}}$, and the user equipment applies a combiner $\mathbf{w}_{m} \in \mathbb{C}^{N_r \times 1}$ from its codebook $\left\{\mathbf{w}_{m}\right\}_{m=1}^{N_{r}}$. The effective received signal can be expressed as $y_{m, n}$. The beam tracking task is to find the optimal beam pair $\tilde{m},\tilde{n}$ at each time slot that maximizes the received signal power $\tilde{m},\tilde{n}=\arg\max_{m, n}|\mathbf{w}_m^H\mathbf{H}(t)\mathbf{f}_n|^2$
with constraints:
\begin{itemize}
    \item Codebook constraint: Both the BS and the UE must select beams from their predefined DFT-based codebooks.
    \item Complexity constraint: The beam tracking must be performed with a limited number of beam measurements, i.e., exhaustive scanning of the full codebook $(N_t \times N_r)$ is impractical in fast-varying channels.
    \item Channel dynamics: The optimization must be repeated over time to cope with temporal variations of the sparse multipath channel $\mathbf{H}(t)$.
\end{itemize}

The beam tracking problem can be expressed as:
\begin{equation}
\begin{aligned}
\underset{m,n}{\text{maximize}} \quad & |\mathbf{w}_m^H\mathbf{H}(t)\mathbf{f}_n|^2 \\
\text{s.t.} \quad & \mathbf{f}_{n} \in \mathbb{C}^{N_t \times 1},\; \mathbf{w}_{m} \in \mathbb{C}^{N_r \times 1},
\label{opt}
\end{aligned}
\end{equation}
This optimization problem is non-convex, and the goal is to estimate the optimal beam pair, which ensures that the transmitter and receiver beams are aligned with the best AoA and AoD. It is crucial to minimize the number of scans while maintaining high accuracy.

\subsection{Channel Charting}

Traditional and hierarchical beam sweeping methods cost too much time and signal overhead, which are inefficient for communications with time-varying channels, especially when there are numerous candidate beams \cite{zhao2024lstm}.

We assume a mobile UE in the radio environment served by a given BS, and we are going to predict beam direction based on the channel chart location of the UE at the served beam. 
Channel charting maps high-dimensional CSI to a low-dimensional chart that reflects the relative spatial distance of the corresponding UEs. Channel chart is constructed from channel features that capture large scale fading effects \cite{studer-vtc}. Therefore, we developed a beam direction prediction strategy based on channel charting, and its mapping relation can be expressed as:
\begin{equation}
    \Phi_{cc} : \{y_{m, n}(t - \tau)\}_{\tau = t_d}^{t_e + t_d - 1} \rightarrow \{\tilde{m}, \tilde{n}\},
\end{equation}
where $t_e$ denotes the time length utilized by the algorithm, which may be variable during execution, and $t_d$ represents the delay of the historical CSI.

Through optimized dimensionality reduction (DR), the radio environment can be mapped in a lower-dimensional space where the relative proximities of UEs are retained. Pairwise feature distance between UEs should be calculated; we use the Euclidean norm to compute the distance between features of UE $k_1$ and UE $k_2$:
\begin{equation}
    \label{euclidean norm}
    d(\mathbf{y}_{k_1},\mathbf{y}_{k_2})=\left\|\mathbf{y}_{k_1}-\mathbf{y}_{k_2}\right\|_2.
\end{equation}
where $\mathbf{y}_k$ is the channel chart feature corresponding to $\mathbf{x}_t$.
We use $d(\mathbf{y}_{k_1},\mathbf{y}_{k_2})$ to measure the similarity of channel chart features in this paper.

The charting function \cite{studer-vtc} is defined as $\mathrm{G}:\mathbb{C}^{M^{^{\prime}}}\to\mathbb{R}^{D^{^{\prime}}}$, i.e.
\begin{equation}
\begin{aligned}
\mathbb{C}^{M^{^{\prime}}}\rightarrow\mathbb{R}^{D^{^{\prime}}} = \mathrm{G}\left(\{y_{m, n}(t - \tau)\}_{\tau = t_d}^{t_e + t_d - 1}\right).
\end{aligned}
\end{equation}

A triplet network architecture is used as DR in this paper; it will be explained in detail later.

For each base station, we construct a corresponding channel chart and annotate its locations with beam direction information. We assume that each UE uses its best beam towards a base station beam. During the offline phase, the beam directions for different BS beams must be measured, requiring the UE to transmit signals across all relevant BS beams. We then train a model and develop a strategy to predict the optimal beam direction based on the channel chart location.

\section{Channel Charting based Beam Tracking}

\begin{figure*}[htp]
    \centering
    \includegraphics[width=17cm]{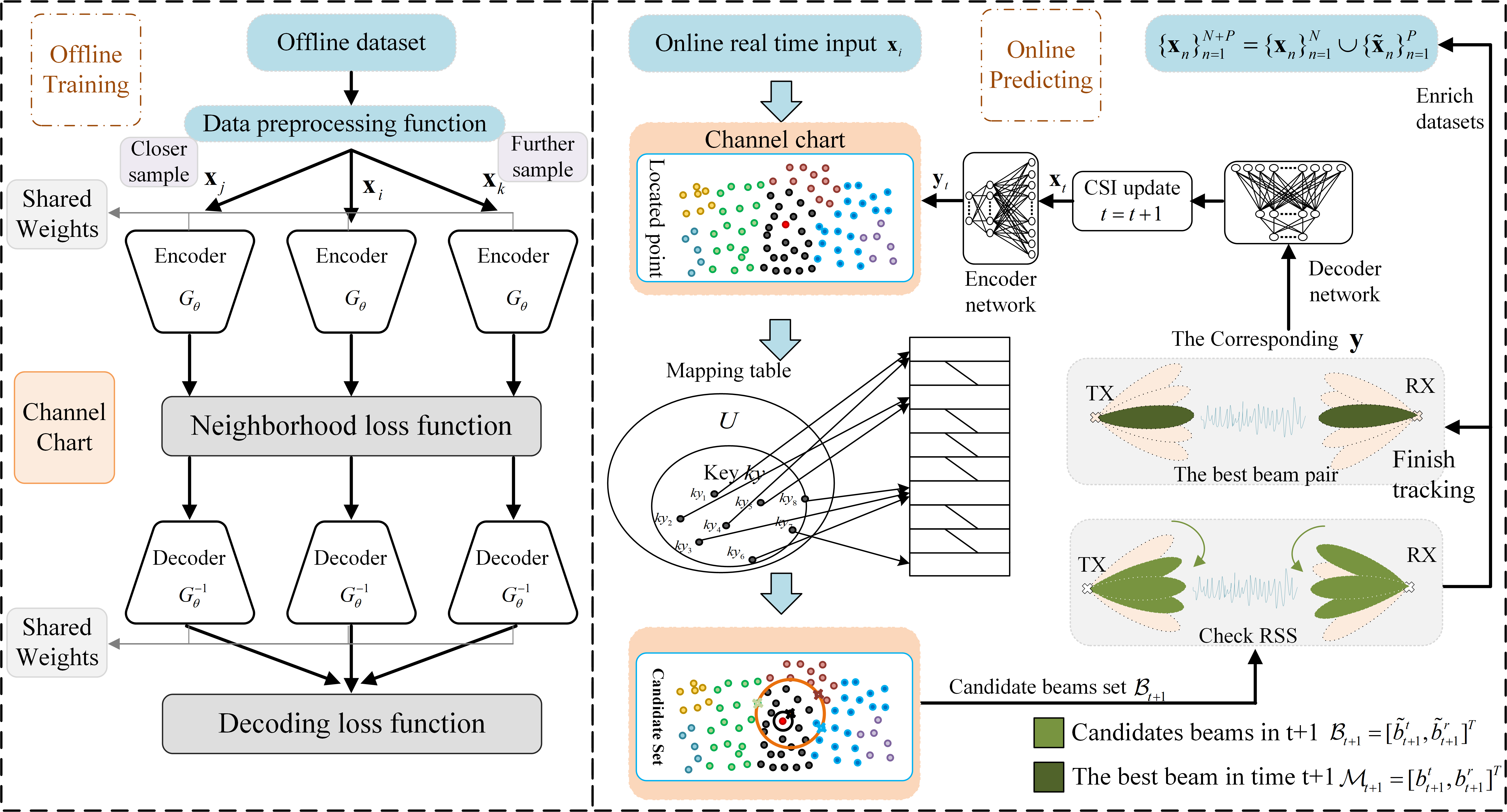}
    \caption{The structure of the channel charting-based beam tracking method including offline training and online predicting.}
    \label{BT_WorkFlow}
\end{figure*}

\subsection{Channel Charting via Neural Network Approximation}

The known Universal Approximation Theorem\cite{hornik1989multilayer} can be stated as follows: Let \( \varphi(\cdot) \) be a bounded and monotonically increasing continuous function, \( \mathcal{I}_d \) a \( d \)-dimensional unit hypercube \( [0,1]^d \), and \( C(\mathcal{I}_d) \) the set of continuous functions defined on \( \mathcal{I}_d \). For any function \( f \in C(\mathcal{I}_d) \), there exists a natural number \( m \) and a set of real numbers \( \nu_i, b_i \in \mathbb{R} \), as well as real vectors \( \mathbf{w}_i \in \mathbb{C}^d \), for \( i = 1, \dots, m \), such that the function \( F \) can be expressed as:
\begin{equation}
F(\mathbf{x}) = \sum_{i=1}^m \nu_i \varphi(\mathbf{w}_i^\mathrm{T} \mathbf{x} + b_i),
\end{equation}
which approximates the function \( f \) as follows:
\begin{equation}
\left| F(\mathbf{x}) - f(\mathbf{x}) \right| < \epsilon, \quad \forall \mathbf{x} \in \mathcal{I}_d,
\end{equation}
where \( \epsilon > 0 \) is a very small number.

In this paper, we use a neural network architecture to approximate the DR function.

Generally, let \( \mathrm{G}(\mathbf{x}_t) \) represent the constructor of the channel graph. It can be expressed as:
\begin{equation}
\mathrm{G}(\mathbf{x}_t) = \left[ F_1(\mathbf{x}_t), \dots, F_k(\mathbf{x}_t) \right],
\end{equation}
where \( F_j(\mathbf{x}_t) = \sum_{i=1}^m \nu_{j,i} \varphi(\mathbf{w}_{j,i}^\mathrm{T} \mathbf{x}_t + b_{j,i}) \), and \( k \) is the dimension of the channel graph. Given the input \( \mathbf{x}_t \), the error can be expressed as follows:
\begin{equation}
\left\| \mathrm{G}(\mathbf{x}_t) - g(\mathbf{x}_t) \right\|_2 < \sqrt{k} \cdot \epsilon.
\end{equation}
where \( g(\mathbf{x}_t) = [f_1(\mathbf{x}_t), f_2(\mathbf{x}_t), \dots]\), and  \( \mathrm{G}(\mathbf{x}_t) \) represent the approximation of \( g(\mathbf{x}_t) \). The composition of $x_t$ will be explained further.

\subsection{Framework}

\begin{figure}[tp] 
	\centering
    \includegraphics[height=7.5cm]{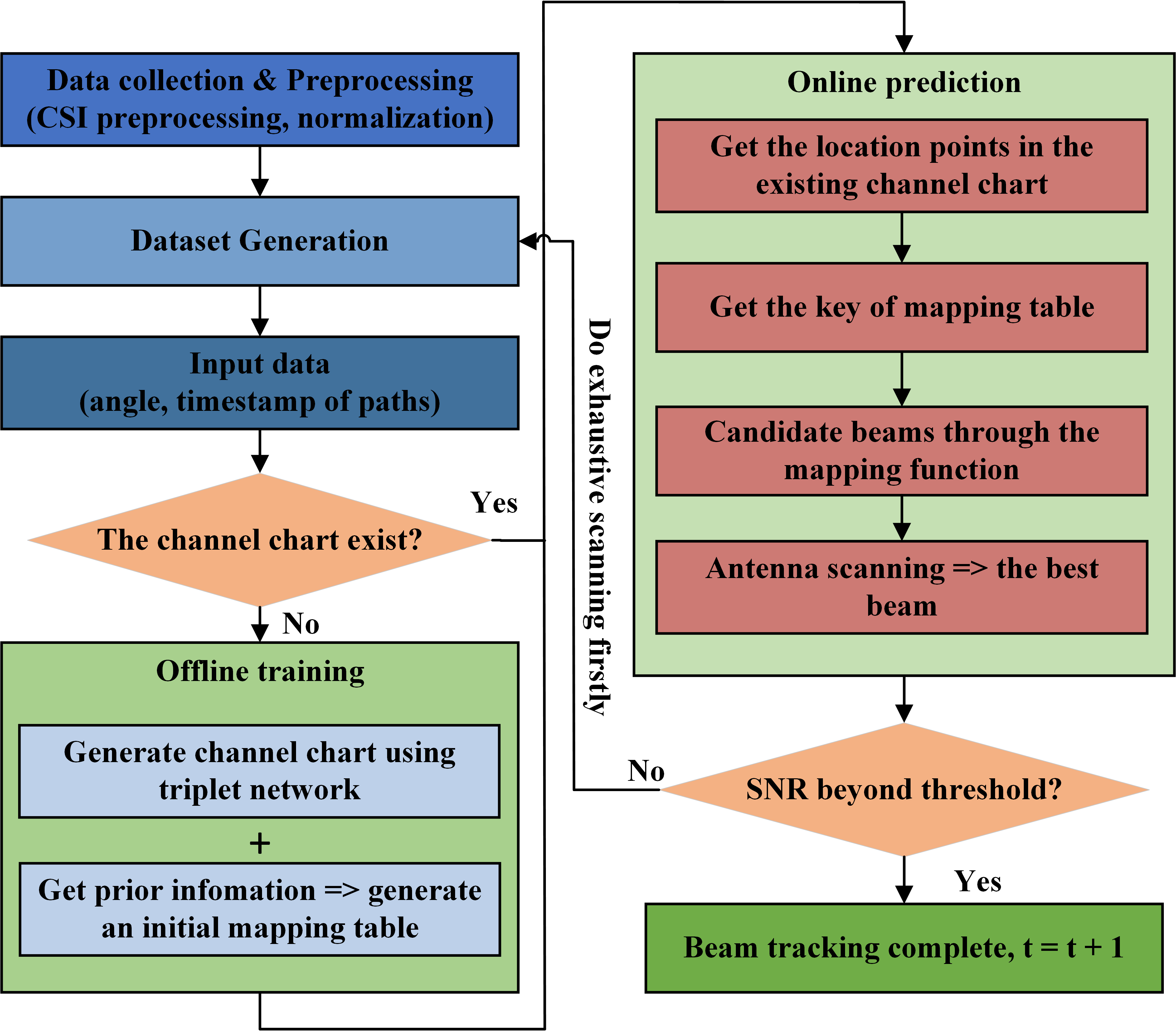}
    \caption{The workflow of the proposed algorithm framework}
    \label{algo-workflow}
\end{figure}

For a static millimeter-wave base station, the proposed channel charting-based beam tracking scheme, illustrated in Fig. \ref{BT_WorkFlow}. The proposed algorithm framework comprises data collection, offline training, online prediction, scanning for confirmation, and data update, as illustrated in Fig. \ref{algo-workflow}. The detailed descriptions are as follows:

\begin{itemize}
    
    \item \textbf{Offline Training}. The data collected by base station from users at different times will be used for training. The data format is given by $\mathbf{x}_t = \left[\alpha_t^{rx},\phi_t^{tx},\alpha_t^{tx},\phi_t^{rx},\tau_t\right]$, where the first four dimensions represent angular information and the last dimension corresponds to temporal information. The parameters $\alpha_t^{rx}$ and $\phi_t^{rx}$ represent the azimuth and elevation of AoA,  $\alpha_t^{tx}$ and $\phi_t^{tx}$ represent the azimuth and elevation of AoD. This training phase employs a triplet network, denoted as $g_{\theta^{}}^{e c}(\cdot)$, and decoder networks, denoted as $g_{\theta^{}}^{d c}(\cdot)$. Channel charting maps high dimensional CSI to low-dimensional chart that reflects relative spatial distance of the corresponding UEs. The current channel information is represented as a feature point, $\mathbf{y}_{t}=g_{\theta}^{ec}\left(\mathbf{x}_{t}\right)$, within the channel chart. During training, triplets of adjacent $\mathbf{x}_t$ are used as input to the network. Meanwhile, a mapping table is constructed based on the mapping function $\mathcal{H}$, it will be explained in detail in Section III.D. 
    
    \item \textbf{Online Predicting}. After communication is established, the BS gives the expected channel charts. The beams prediction process is described in Fig. \ref{BT_WorkFlow}. Firstly, the neighborhood search is performed to get the initial beam. The current channel angle feature is set as $\mathbf{x}_{t}$, and the corresponding channel latent feature is $\mathbf{y}_{t}$. We define $\mathbf{y}_{t}$ as the location point of $\mathbf{x}_{t}$, which is dotted red on the left side of Fig.\ref{BT_WorkFlow}. We use the Breadth First Search (BFS) algorithm to get the mapped point because BFS will get the first label point it touches. We assume the closest label should represent  $\mathbf{x}_{t}$'s beam feature. Secondly, we get its neighbor beams. Note that the neighbor beam index should be between 1 and the total number of beams. Thus, modular arithmetic is needed. Meanwhile, we need to take out the candidate beams it would change abruptly from the mapping table, which is recorded in the training phase. Finally, a small number of candidate beams are chosen as the next scanning beam set $\mathcal{B}_{t+1} $.
    
    \item \textbf{Scanning for confirmation}. According to the given detection range $\mathcal{B}_{t+1}$, a minimal amount of beam scanning is performed to determine the best transmission beam, which also provides information for the next prediction. While reaching the time interval for triggering beam tracking, both the BS and the UE scan the predicted beam set $\mathcal{B}_{t+1}$ to obtain the currently best beam pair indexes $\mathcal{W}_{t+1}=\left[b_{t+1}^{t}, b_{t+1}^{r}\right]^{T}$. Then the channel angle information $\mathbf{x}_{t+1}=\left[\alpha_{t+1}^{rx},\phi_{t+1}^{tx},\alpha_{t+1}^{tx},\phi_{t+1}^{rx},\tau_{t+1}\right]^{T}$ can be obtained according to the trained decoder \(G_{\theta}^{-1}\) to reconstruct a CSI feature vector, because auto-encoders learn bijective manifolds\cite{auto-encoder}. Specifically, after obtaining the latent coordinate \(y_t\), the BS/UE computes 
    $\hat{x}_{t+1} = G_{\theta}^{-1}\!\left(\mathbf{y}_{t}\right)$
    The four angle components are then quantized to the nearest DFT-codebook indices, yielding the integer vector $\mathbf{x}_{t+1}=\left[\alpha_{t+1}^{rx},\phi_{t+1}^{tx},\alpha_{t+1}^{tx},\phi_{t+1}^{rx},\tau_{t+1}\right]^{T}$. Repeat the above process, and robust beam tracking can be achieved.

    \item \textbf{Data Update}. The final step is to update our database. When the beam we get cannot satisfy the needs of communication, we will do an exhaustive search to get the current best beam. Meanwhile, this information will be connected to the corresponding $\mathbf{y_i}$. Then, the next time we meet $\mathbf{y_i}$, our algorithm will consider this beam as one of the candidate beams, with high priority. The updated database provides an increasingly accurate beam tracking service over time.
    
\end{itemize}

The prediction step of our algorithm generates the codebook index for the next time step directly from the model $\mathcal{B}_{t+1}$, based on the output of the previous time step. Initially, the input data undergoes compression, followed by the extraction of two-dimensional features through the encoding module, get $\mathbf{y}_t$. Subsequently, the candidate beam set for the next time step is rapidly determined with the assistance of a mapping table.

After obtaining the candidate beams, the antenna scans and chooses the direction with the highest SNR as the predicted beam direction. If the SNR does not meet the threshold, beam tracking is considered a failure, and the antenna should perform an exhaustive scan for beam alignment. Meanwhile, the information from this attempt is recorded to improve the framework's robustness.

The proposed algorithm supports multipath scenarios by directly aligning the output direction with the antenna's codebook. Using deep learning to extract features from CSI, the algorithm clusters CSI from different paths in the channel chart, effectively capturing similarities and differences in multipath channels. 

\subsection{CSI Preprocessing}

One of the most critical aspects in the design of good features for CC is to realize that CSI in radi geometry is a poor representation of spatial geometry \cite{studer_channel_2018}. To overcome this challenge, CSI data need to be preprocessed.

The raw $2^{nd}$ moment (R2M) data of the channel is necessary in generating the channel chart, we can simplify the computation of $\mathbb{E}\left[\mathbf{h}_t\mathbf{h}_t^H\right]$ to the computation of $\mathbb{E}\left[\mathbf{x}_t\mathbf{x}_t^H\right]$\cite{studer_channel_2018}. The R2M in this paper can be expressed as follows:
\begin{equation}
\mathbf{\bar{X}}=\mathbb{E}\big[\mathbf{x}\mathbf{x}^H\big]=\frac1T\sum_{t=1}^T\mathbf{x}_t\mathbf{x}_t^H.
\label{csi_scaling}
\end{equation}

In practical operations, we obtain it by averaging the CSI collected over \(T\) short time intervals.

We use the Euclidean norm  as a dissimilarity measure for pairs of features, as shown in (\ref{euclidean norm}). In wireless environments, due to path loss, the geometric spatial characteristics of CSI are relatively poor. For example, $Rx_C$ and $Rx_D$ are located further from the transmitter, and their dissimilarity is greater than that of $Rx_A$ and $Rx_B$. But $d(\mathbf{X}_{C},\mathbf{X}_{D})<d(\mathbf{X}_{A},\mathbf{X}_{B})$ due to the path loss. Therefore, the R2M obtained needs to be scaled \cite{studer_channel_2018}:

\begin{equation}
\tilde{\mathbf{X}}=\frac{\Omega^{\beta-1}}{\left\|\bar{\mathbf{X}}\right\|_F^\beta}\bar{\mathbf{X}},\beta=1+1/(2\sigma).
\label{csi-scaling-end}
\end{equation}
where $\Omega$ is the path loss factor, and $\sigma\in[0,\infty]$ is a path loss parameter. Due to $\beta\geq1$, the CSI distant from the transmitter is amplified, while the CSI close to the transmitter is attenuated.

\subsection{Channel Charting Model}
We construct an autoencoder using a multilayer neural network, enabling the autoencoder function to map high-dimensional channel data into a two-dimensional space. This process establishes a mapping relationship between the CSI and coordinates within the channel chart. The concept of the autoencoder leads to the training of two functions: the encoding function $\mathrm{G}:\mathbb{C}^{M^{^{\prime}}}\to\mathbb{R}^{D^{^{\prime}}}$ and the decoding function $\mathrm{G}^{-1}{:}\mathbb{R}^{D^{\prime}}\to\mathbb{C}^{M^{\prime}}$, $M'>D'$, the error can be represented as:

\begin{equation}
E(\mathbf{x}_n)=\frac1N\sum_{n=1}^N\parallel\mathbf{x}_n-\mathrm{G}^{-1}(\mathrm{G}(\mathbf{x}_n))\parallel_2^2.
\label{rec_e}
\end{equation}

In this paper, $M'$ represents the dimensionality of the CSI, whereas $D'$ denotes the spatial dimension of the channel chart. The reconstruction error of the input vector should be minimized, that is, $\mathbf{x}_{n}\approx\mathrm{G}(\mathrm{G}^{-1}(\mathbf{x}_{n}))$.

In practice, the dimensionality reduction performance of multilayer networks is generally superior to that of single-layer networks. For each layer within the autoencoder, its operation can be considered to involve multiplying the input by a weight matrix and adding a bias term. Under this situation, $\mathrm{G}$ and $\mathrm{G}^{-1}$ can be specifically represented as:
\begin{equation}
\begin{aligned}
&\mathrm{G}:\mathbf{y}=g_{ec}\left(\mathbf{x}\right)=\mathbf{W}_{ec}\mathbf{x}+\mathbf{b}_{ec}, \\
&\mathrm{G}^{-1}:\mathbf{\hat{x}}=g_{dc}\left(\mathbf{y}\right)=\mathbf{W}_{dc}\mathbf{y}+\mathbf{b}_{dc},
\end{aligned}
\label{net-model}
\end{equation}
where $g_{ec}$ and $g_{dc}$ are encoding function and decoding function, $\mathbf{W}$ and $\mathbf{b}$ represent the weight matrix and adding the bias. $\mathbf{x} \in \mathbb{C}^{d_x}, \mathbf{y} \in \mathbb{R}^{d_y}$, with $d_x = M', d_y = D', \text{and} \quad M' > D'$. $\mathbf{W}_{ec}\in\mathbb{C}^{d_y\times d_x},\quad \mathbf{b}_{ec}\in\mathbb{R}^{d_y}$ and $\mathbf{W}_{dc}\in\mathbb{R}^{d_x\times d_y},\quad \mathbf{b}_{dc}\in\mathbb{C}^{d_x}$. This channel charting method only involves matrix operations, which are convenient to implement in practical systems.

After constructing a single network model, inspired by the literature \cite{schroff_facenet_2015}, we introduce a triplet network architecture to generate the channel chart. This architecture enables the maintenance of relative positional relationships within the actual space.

Assuming the channel information vector is represented by $\mathbf{x}$, according to (\ref{csi_scaling}), R2M can be calculated, which is then flattened into a vector to serve as the input to the network. The training data for each input exists in the form of triplets $[\mathbf{x}_i,\mathbf{x}_j,\mathbf{x}_k]$, and their interrelations satisfy the following conditions:
\begin{equation}
d(\mathbf{x}_i,\mathbf{x}_j)\leq d(\mathbf{x}_i,\mathbf{x}_k).
\end{equation}
where $\mathbf{x}_i$ is called the anchor point, $\mathbf{x}_j$ is called the positive point, and $\mathbf{x}_k$ is called the negative point.

Taking inspiration from the design of the loss function in the literature \cite{ferrand_triplet-based_2021}, the design of the neighborhood loss error of our triplet network architecture can be represented as:
\begin{equation}
\mathcal{C}_{n}=\frac{1}{\left|\mathcal{T}\right|}\sum_{(i,j,k)\in\mathcal{T}}\left(
\begin{aligned}
\left\|\mathrm{G}_{\theta}(\mathbf{x}_{i})-\mathrm{G}_{\theta}(\mathbf{x}_{j})\right\|- \\ \left\|\mathrm{G}_{\theta}(\mathbf{x}_{i})-\mathrm{G}_{\theta}(\mathbf{x}_{k})\right\|
\end{aligned}
\right)^{+},
\label{cost1}
\end{equation}
where $\mathcal{T}$ represents the total number of training samples. $\begin{pmatrix}x\end{pmatrix}^+$ is the abbreviation of $\max(x,0)$.

In order to increase the distance between negative sample points and the anchor point, a constant needs to be added to the neighborhood error function, thereby preserving the spatial relationships of the channel features:

\begin{equation}
\mathcal{C}_{n}=\frac{1}{\left|\mathcal{T}\right|}\sum_{(i,j,k)\in\mathcal{T}}\left(
\begin{aligned}
\left\|\mathrm{G}_{\theta}(\mathbf{x}_{i})-\mathrm{G}_{\theta}(\mathbf{x}_{j})\right\|- \\ \left\|\mathrm{G}_{\theta}(\mathbf{x}_{i})-\mathrm{G}_{\theta}(\mathbf{x}_{k})\right\|+  \Delta
\end{aligned}
\right)^{+}.
\label{cost2}
\end{equation}

This makes sure that $\left\|\mathrm{G}_{\theta}(\mathbf{x}_{i})-\mathrm{G}_{\theta}(\mathbf{x}_{j})\right\|+ \Delta>\left\|\mathrm{G}_{\theta}(\mathbf{x}_{i})-\mathrm{G}_{\theta}(\mathbf{x}_{k})\right\|$. The triplet loss in (\ref{cost2}) is adopted to preserve the spatial topology of the channel features in the latent chart domain.

Note that the cost function in (\ref{cost2}) has a trivial solution,
which maps all samples to the same point in Y and achieves zero cost \cite{ferrand_triplet-based_2021}. In order to avoid this uninteresting solution, some slack can be added to the cost function through $\Delta$. In our implementation, $\Delta$ can be chosen as 1 without loss of generality \cite{ferrand_triplet-based_2021}.

The network's decoding loss is shown as in (\ref{rec_e}), define $\mathcal{C}_{r}(\mathbf{x}_n)=E(\mathbf{x}_n)$. The results of network training can be regarded as the solution to the following optimization problem:
\begin{equation}
\tilde{\mathrm{\theta}}=\arg\min_{\mathrm{\theta}}\left(\mathcal{C}_{n}+\mathcal{C}_{r}\right).
\end{equation}

Based on the channel chart mapping function obtained from training with DNN, once the parameters $\tilde\theta$ are determined, the channel chart coordinates of any new input vector $\mathbf{x}$ can be obtained by calculating the value of the function $\mathrm{G}_{\tilde\theta}(\mathbf{x})$.



When the area of interest changes, there is no need for complete offline retraining. Instead, the channel charting can be updated through incremental learning and local adjustments. When new users join, the system can leverage the existing channel charting to obtain effective candidate beams, while adapting to the new environment by updating the mapping relationships in the local regions. The algorithm is highly practical because it avoids frequent full retraining and can efficiently respond to environmental changes through incremental updates and adaptive mechanisms. However, the timeliness of the channel charting is carefully considered. Later sections will determine the retraining cycle of the algorithm based on the fluctuations in the prediction accuracy.

\subsection{Obtain Candidate Beam Set}
\begin{figure*}[tp]
    \centering
    \includegraphics[height=5.8cm]{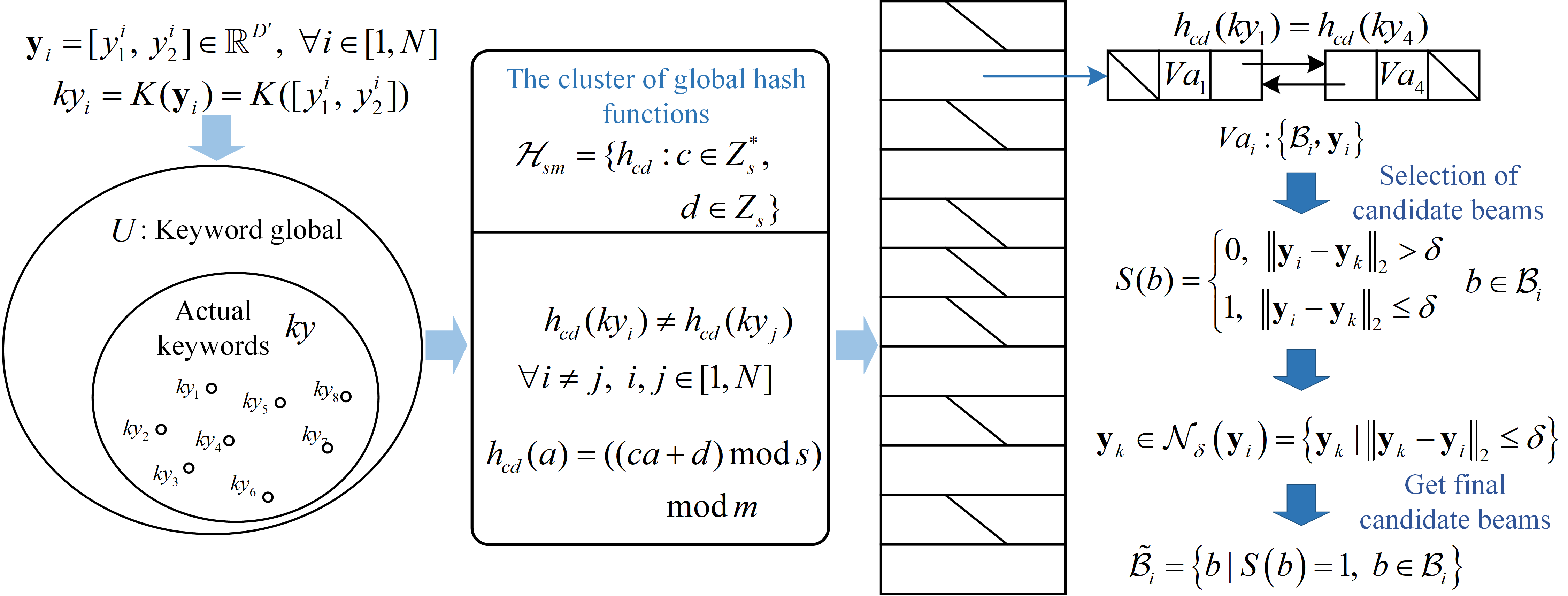}
    \caption{The strategy that we constructe to obtain the set of candidate beams based on a hash table.}
    \label{mapping-table}
\end{figure*}

In our previous work, we utilized the Breadth-First Search (BFS) algorithm twice to generate two candidate beams \cite{our-pre-work}. In this paper, we propose an enhancement to this approach by incorporating a mapping table. The core idea behind this enhancement is to establish a mapping between each point in the channel chart and the corresponding candidate beam set, thereby enabling immediate retrieval of candidate results when required. The mapping table is implemented as a hash table, which allows for its analysis within the theoretical framework of hash table structures.

The framework for obtaining candidate beams after channel charting is shown in Fig. \ref{mapping-table}. First, the algorithm generates a key based on the positioning results of the channel chart. This key is then mapped to a specific slot through a mapping function, from which the corresponding set of candidate beams is retrieved. Finally, after a filtering process, the antenna scans the filtered candidate beams to complete the current beam-tracking task.

The expected performance of the mapping table depends on the chosen mapping function \(h\), affecting the query time for the candidate beam sets. Ideally, the distribution of keys mapped across the table should be uniform. There exists a mapping function \(h\) that maps the entire domain of keys to the slots in the mapping table, as follows:
\begin{equation}
h:U\to\{0,1,...,m-1\},
\end{equation}

where $U$ is the universal set, \(m\) is generally much smaller than $\left|U\right|$, and \(h(ky)\) is referred to as the hash value of the key \(ky\).

The length of any linked list \(T_i\) on the mapping table can be represented by \(n_i\), thus $n=\sum_{i=1}^mn_i$. The expected value of \(n_i\) is $\mathbb{E}[n_{i}]=n/m$. For any coordinate $\mathbf{y}_{i}=[y_{1}^{i},y_{2}^{i}]$ in the channel diagram, define the key generation function as \(K\):
\begin{equation}
ky=K(\mathbf{y}_i)=c^{k+1}\left(\left\lfloor k\cdot y_1\right\rfloor+\left\lfloor k\cdot y_2\right\rfloor\right) ,\end{equation}
where $c$ is a generating factor, which can be a constant, $\lfloor\cdot\rfloor $ is the floor function operator, \(k\) depends on the length of $ky$. Therefore, the generated set of keys can be represented as:
\begin{equation}
\begin{Bmatrix}ky_n\end{Bmatrix}_{n=1}^N=K\biggl(\bigl\{\mathbf{y}_i\bigr\}_{i=1}^N\biggr),\mathbf{y}_i=\bigl[y_1^i,y_2^i\bigr].
\end{equation}

Through key mapping, the candidate beam sets can be obtained, $\left\{\mathcal{B}_n\right\}_{n=1}^N\overset{h(\cdot)}{\leftarrow}\left\{ky_n\right\}_{n=1}^N$. The candidate sets obtained vary with different mapping strategies. 

We employ the chaining method to resolve collision issues in the mapping table, and it can be demonstrated that the query time complexity for the candidate beam sets is $\mathcal{O}(1+\alpha)$. 

Assuming that the mapping table holds \(n\) candidate beam sets across \(m\) slots, and that the element \(e\) to be queried could be any one of them, with equal probability, then the number of elements accessed to query \(e\) is the number of elements before \(e\) plus 1. Based on this, the expected complexity for each query can be calculated. Assuming \(e_i\) is the \(i\)th candidate set inserted into the table, and assuming \(k_i\) is the \(i\)th key, define the random variable \(X_{ij}=I\{h(k_i)=h(k_j)\}\), where \(i\) is the indicator function. Consequently, \(\mathbb{E}[X_{ij}]=\frac{1}{m}\) indicates the expected probability that two keys hash to the same slot. Therefore, the time complexity for a single query can be expressed as \cite{introduction_to_algo}:

\begin{equation}
\begin{aligned}
\mathbb{E}\Big[\frac{1}{n}\sum_{i=1}^{n}\Bigg(1+\sum_{j=i+1}^{n}X_{ij}\Bigg)\Big]=\frac{1}{n}\sum_{i=1}^{n}\Bigg(1+\sum_{j=i+1}^{n}\mathbb{E}[X_{ij}]\Bigg) \\
=1+\frac{1}{nm}\sum_{i=1}^{n}(n-i)
=1+\frac1{mn}\Big(n^2-\frac{n(n+1)}2\Big),
\end{aligned}
\end{equation}

Let $\alpha=n/m$, the above equation can be further expressed as:
\begin{equation}
\begin{aligned}
\mathbb{E}\Big[\frac{1}{n}\sum_{i=1}^{n}\Bigg(1+\sum_{j=i+1}^{n}X_{ij}\Bigg)\Big]
&=1+\frac{n-1}{2m} \\
&=1+\frac\alpha2-\frac\alpha{2n}.
\end{aligned}
\end{equation}
For large $n$, the term $\frac{\alpha}{2n}$ becomes negligible.

Thus, the time complexity cost of obtaining candidate beam sets from the channel diagram coordinates is \(\mathcal{O}(1 + \alpha)\), where \(\alpha\) is the load factor, representing the average number of elements per slot in the hash table. If \(m\) is proportional to \(n\), i.e., \(n = \mathcal{O}(m)\), then $\alpha=n/m=\mathcal{O}(m)/m=\mathcal{O}(1)$.

Universal hashing functions exhibit good average performance. Compared to ordinary mapping functions, they have a lower probability of collisions and do not require consideration of the specific distribution of keys. During execution, universal mapping randomly selects a set of well-designed functions, resulting in different results for each execution of the algorithm. Assuming $\mathcal{H}$ represents a set of universal hash functions, for any pair of distinct keys \(a\), \(b\), the number of hash functions \(h\) for which \(h(a) = h(b)\) is at most $\mathcal{H}/m$. This implies that, when \(a \neq b\), $\Pr\{h_\theta(a)=h_\theta(b)\}\leq1/m$. For any one event $A$, $\mathbb{E}[X_A]=\operatorname{Pr}\{A\}$, so we can get $\mathbb{E}[X_{ab}]\leq1/m$.

For each keyword \(a\) generated by the channel chart, define the random variable \(Y_a'\) to represent the number of other keywords that are assigned in the same slot as \(a\). It follows that
\begin{equation}
Y_a^{\prime}=\sum\limits_{\substack{b\neq a\\b\in T}}X_{ab}
\end{equation}
and
\begin{equation}
\mathbb{E}[Y_a']=\mathbb{E}\Bigg[\sum\limits_{\substack{b\neq a\\b\in T} }X_{ab}\Bigg]=\sum\limits_{\substack{b\neq a\\b\in T}}\mathbb{E}[X_{ab}]\leq\sum\limits_{\substack{b\neq a\\b\in T}}\frac{1}{m}.
\end{equation}

The remaining parts can be discussed by category:
\begin{enumerate}
    \item If \( a \notin T\), that is, the new keyword generated by the channel chart coordinates is not present in the mapping table, then $n_{h_\theta(a)} = Y_a'$, and $\begin{vmatrix}b:b\in T,b\neq a\end{vmatrix}=n$, so $\mathbb{E}[n_{h_{\theta}(a)}]{=}\mathbb{E}[Y_{a}^{\prime}]{\leq}n/m {=}\alpha $.
    \item If \( a \in T\), the keyword \( a \) is present in the linked list \( T \) and $a \notin Y_a'$, so $n_{h_\theta(a)}=Y_a'+1$, $\begin{vmatrix}b:b\in T,b\neq a\end{vmatrix}=n-1$, we get $\mathbb{E}[n_{h_{\theta}(a)}]=\mathbb{E}[Y_{a}^{\prime}]+1 \leq(n-1)/m+1=a+\alpha-1/m<1+\alpha$.
\end{enumerate}

Based on knowledge of number theory, a universal mapping function can be easily constructed. First, let \(s\) be a sufficiently large prime number, ensuring that every possible key falls within the interval from 0 to \(s-1\). Assume \( Z_s \) represents the set \( \{0, 1, \dots, s-1\} \), and \( Z_{s}^{*} \) denotes the set \( \{1, 2, \dots, s\} \). Since the total number of keywords exceeds the number of slots, it follows that \( s > m \). For any $c\in Z_s^*$ and $d\in Z_s$, we define the mapping function $h_{cd}$: $Z_s \rightarrow Z_m$ as:
\begin{equation}
h_{cd}(a)=((ca+d)\operatorname{mod}s)\operatorname{mod}m,
\end{equation}
where $a \in Z_s$ is the key to be mapped. For example, when $s=18,m=7$, $h_{3,4}(8)=3$.

The family of hash functions, comprising all functions of the form defined above, is represented as:
\begin{equation}
\mathcal{H}_{sm}=\{h_{cd}:c\in Z_s^*,d\in Z_s\}.
\label{quanyu-hanshu}
\end{equation}

Each \( h_{cd} \) maps elements from \( Z_s \) onto \( Z_m \).

Under the influence of $\mathcal{H}_{sm}$, $y_j$ can map to the collection of all candidate beam sets on a chain in the mapping table, and a specific set can be determined based on the keyword $ky$:

\begin{equation}
\mathbf{y}_j\xrightarrow{K}ky_j\xrightarrow{h_{cd}}T_{h_{cd}(ky_j)}\xrightarrow{\mathrm{Search}}\mathcal{B}_j.
\label{get-beam-set}
\end{equation}

Instead of directly predicting beam probabilities using a neural network, a hash table is constructed to associate each latent feature vector $\mathbf{y}$ with a small set of candidate beams. The hash key is generated from the discretized latent coordinates, and the value stores the beam indices most frequently observed in that latent region.

This design converts the beam search process into a constant-time lookup operation and supports incremental updates when the environment or user trajectory changes.

\subsection{Selection of candidate beams}
As generating candidate beams, due to numerical precision issues, identical keywords generated by different $\mathbf{y}_j$ values result in candidate beams that belong to the same set. In this scenario, the number of antenna scans required will increase significantly.

We leverage the characteristics of CC to involve the selection rules for candidate beams. For any slot \( T_i \) in the mapping table and its associated set of candidate beams \( \mathcal{B}_i \). The set of positions in the channel chart corresponding to the beams in \( \mathcal{B}_i \) is $\{\mathbf{y}_k\mid b_k\in\mathcal{B}_i\}$. The \(\delta\)-neighborhood of a point $\mathbf{y_i}$ is defined as:
\begin{equation}
\mathcal{N}_\delta(\mathbf{y}_i) = \left\{ \mathbf{y}_k \mid \|\mathbf{y}_i - \mathbf{y}_k\|_2 \leq \delta \right\},
\end{equation}
where $\mathbf{y_k}$ are positions in the channel chart.
The range of \(\delta\) is $[\delta_{\min},\delta_{\max}] = [\min\left(\left\|\mathbf{y}_i-\mathbf{y}_j\right\|_2\right), \max\left(\left\|\mathbf{y}_i-\mathbf{y}_j\right\|_2\right)]$,
$\mathbf{y}_i$ is any point in the channel chart, and \(\mathbf{y}_j \) is the closest point to \( \mathbf{y}_i \). In practice, the value of \(\delta\) can be determined within its range using the binary search method.

Therefore, for the candidate beam set \( \mathcal{B}_i \) associated with the slot \( T_i \) in the mapping table, the selection criteria for each candidate beam $b\in\mathcal{B}_i$ can be defined as:
\begin{equation}
\mathcal{S}(b)=\begin{cases}0,&\left\|\mathbf{y}_i-\mathbf{y}_k\right\|_2>\delta\\1,&\left\|\mathbf{y}_i-\mathbf{y}_k\right\|_2\leq\delta\end{cases}.
\end{equation}
where $\mathbf{y}_i$ is the query point in the channel chart (the current position), $\mathbf{y}_k$ is the position corresponding to candidate beam $b$.

The function $\mathcal{S}(b)$ indicates whether to include the candidate beam $b$ in the final set based on its proximity to $\mathbf{y_i}$. If the distance between $\mathbf{y_i}$ and $\mathbf{y_k}$ is within $\delta$, we select $b$; otherwise, we discard it. As a result, the final beam set is $\tilde{\mathcal{B}_i}=\{b\mid\mathcal{S}(b)=1, b\in\mathcal{B}_i\}$.

The algorithm presented in this paper is designed to be compatible with the 5G NR communication standard. It functions by conducting antenna sweeping on both the base station and user equipment sides, utilizing synchronization signal blocks. Beam tracking is typically performed at intervals ranging from 20 to 100 ms, ensuring both reliability and communication efficiency in highly dynamic environments.

\section{The Beam Tracking Process And Error definition}
In this section, we will show how we will run the proposed algorithm and the two types of evaluation errors: positioning error and prediction error.
\subsection{Tracking Process}
Assuming that the channel chart $\mathcal{F}_{t}$ and $\mathcal{F}_{r}$ have already been constructed. Firstly, in global timing, for each new input \( x_t \), the positional coordinates of the channel chart $\mathbf{y}_t$ are obtained through the mapping function \( \mathrm{G}_{\theta} \). Further, algorithm will generate the set of beam candidates \( \mathcal{B}_i \) for each input instance. Eventually, by scanning each direction with the antenna, the index of the best beam is recorded, leading to the determination of the best beam index pair $\mathbf{b}_{t+1}=[I_{t+1}^t,I_{t+1}^r]$ for the next instance. During the process, the input \( \mathbf{x}_{n} \) is processed through the decoding network \( \mathrm{G}_{\theta}^{-1} \) to compute the overall network error \( \mathcal{C}_{r} \), which assesses the efficacy of the generated channel chart network. By acquiring the channel information vector of the optimal beam and updating the temporal information to get \( \mathbf{x}_{t+1} \), this process is repeated continuously; we can achieve real-time beam tracking. The above description is summarized in Algorithm \ref{alg:alg2}.

\begin{algorithm}[t]

\caption{Beam Tracking Process}\label{alg:alg2}
\begin{algorithmic}
    \State \textbf{Input:} $\mathbf{x}_{t}$, $\mathrm{G}_{\theta}$, $\mathrm{G}_\theta^{-1}$, mapping table $T$.
    \State \textbf{Output:} Best beam pair $\mathbf{b}_{t+1}=[b_{t+1}^t,b_{t+1}^r]$ at $t+1$.
    \State\textbf{Offline training:} $\mathbb{C}^{M^{^{\prime}}}\rightarrow\mathbb{R}^{D^{^{\prime}}}$, i.e., $\mathbf{x_t}\rightarrow \mathbf{y_t}$
    \State\textbf{Tracking Process: $\{y_{m, n}(t - \tau)\}_{\tau = t_d}^{t_e + t_d - 1} \rightarrow \{\tilde{m},\tilde{n}\}$}
    \State\hspace{0.5cm} (1) Generate the trained channel charting DR function $\mathrm{G}_{\theta}$.
    \State\hspace{0.5cm} (2) Feed $\mathbf{x}_t = \left[\alpha_t^{rx},\phi_t^{tx},\alpha_t^{tx},\phi_t^{rx},\tau_t\right]$ to $\mathrm{G}_{\theta}$, obtain the 
low-dimensional representation $\mathbf{y}_t$ 
\State\hspace{1.3cm}by neighborhood search algorithm.
    \State\hspace{0.5cm} (3) Feed $\mathbf{y}_t$ to $h_{cd}$, obtain the candidate beam sets $\mathcal{B}_t^{tx}$ 
and $\mathcal{B}_t^{rx}$.
    \State\hspace{0.5cm} (4) Scan for confirmation, get the next beam pair $\mathbf{b}_{t+1}$.
    \State\hspace{0.5cm} (5) Check the SNR of $\mathbf{b}_{t+1}$. If $\mathbf{b}_{t+1}$ not in $[\mathcal{B}_t^{tx}, \mathcal{B}_t^{rx}]$, do misalignment handling.
    \State\hspace{0.5cm} (6) Calculate loss $\mathcal{C}_{r}$, update CSI for $\mathbf{x}_{t+1}$.
    \State\hspace{0.5cm} (7) Let $t+1 \rightarrow t$.
    \State\textbf{Misalignment Handling}
    \State\hspace{0.5cm} \textbf{Loop each $\mathbf{b}\in[\mathcal{B}_t^{tx}, \mathcal{B}_t^{rx}]$}:
    \State\hspace{0.5cm} (1) Calculate the $SNR$ of beam pair $\mathbf{b}$.
    \State\hspace{0.5cm} (2) \textbf{if} $SNR$ not meet the threshold
    \State\hspace{1.5cm} accumulate $cnt_{ms}$.
    \State\hspace{0.5cm} (3) \textbf{if} $cnt_{ms}=len\left([\mathcal{B}_t^{tx}, \mathcal{B}_t^{rx}]\right)$
    \State\hspace{1.5cm} Exhaustive search, record the best beam $\mathbf{b_m}$.
    \State\hspace{0.5cm} Feedback $\mathbf{b_m}$, and add it into training dataset.
\end{algorithmic}

\label{alg2}
\end{algorithm}

\subsection{Error Definition}
In this paper, two indicators are used to evaluate the accuracy of the algorithm: positioning error and prediction error. Positioning errors in the algorithm do not necessarily result in prediction errors. 

The positioning error is defined as:
\begin{equation}
\begin{aligned}
E_{ps}=E_{ps}^{t}+E_{pos}^{r}=\sum_{n=1}^{N_{\text{alg}}}\mathbb{I}_{b^{t}(n)\neq\hat{b}^{t}(n)}+\sum_{n=1}^{N_{\text{alg}}}\mathbb{I}_{b^{r}(n)\neq\hat{b}^{r}(n)},
\end{aligned}
\end{equation}
where $N_{\text{alg}}$ is the total number of algorithms run, $b^{t}(n),b^{r}(n)$ are the beam indices of both sides' located points at time $n$, $\hat{b}^{t}(n),\hat{b}^{r}(n)$ are both sides' best beam indices. $\mathbb{I}_{b^{t}(n)\neq\hat{b}^{t}(n)}$ is defined as 1 if the condition is met, and 0 otherwise.

When scanning the candidate beam set, if the best communication beam is included in $\mathcal{B}_i$, then the antenna can capture it. We define the indicator function \( e(n) \) at moment \( n \); if the best communication beam does not fall within \( \mathcal{B}_n \), \( e(n) \) is set to 1; otherwise, it is set to 0. Therefore, the total prediction error can be defined as:
\begin{equation}
E_{pd}=E_{pd}^t+E_{pd}^r=\sum_{n=1}^Ne^t(n)+\sum_{n=1}^Ne^r(n).
\end{equation}

In practice, when the positioning operation takes too long, we employ the neighborhood search method proposed in study \cite{our-pre-work} to obtain two candidate beams, ensuring the real-time performance of the algorithm. In this context, the total number of searches can be expressed as:
\begin{equation}
N_{s}=\sum_{n=1}^{N-\zeta^{(t)}}\left|\mathcal{B}^{tx}_n\right|+2\cdot \zeta^{(t)}+\sum_{n=1}^{N-\zeta^{(r)}}\left|\mathcal{B}^{rx}_n\right|+2\cdot \zeta^{(r)}.
\end{equation}
where \(\zeta^{(t)}\) and \(\zeta^{(r)}\) respectively represent the number of neighborhood searches at both the transmitter and the receiver during computation. $\mathcal{B}^{tx}_n$ and $\mathcal{B}^{rx}_n$ denote the number of candidate beams at time $n$.

\section{Experiments and Discussions}
In this section, we have considered four experimental subsections to evaluate the performance of the algorithm proposed in this paper. These include different configuration parameters of the algorithm, comparative accuracy of various algorithms, timeliness analysis of the algorithm, and actual field testing. Data in the simulation environment are derived from a ray tracing model, while the field testing data are collected from an actual 26 GHz millimeter-wave communication platform.




\subsection{Runtime and comparison}

\begin{figure}[tp]
    \centering
    \includegraphics[width=8.5cm]{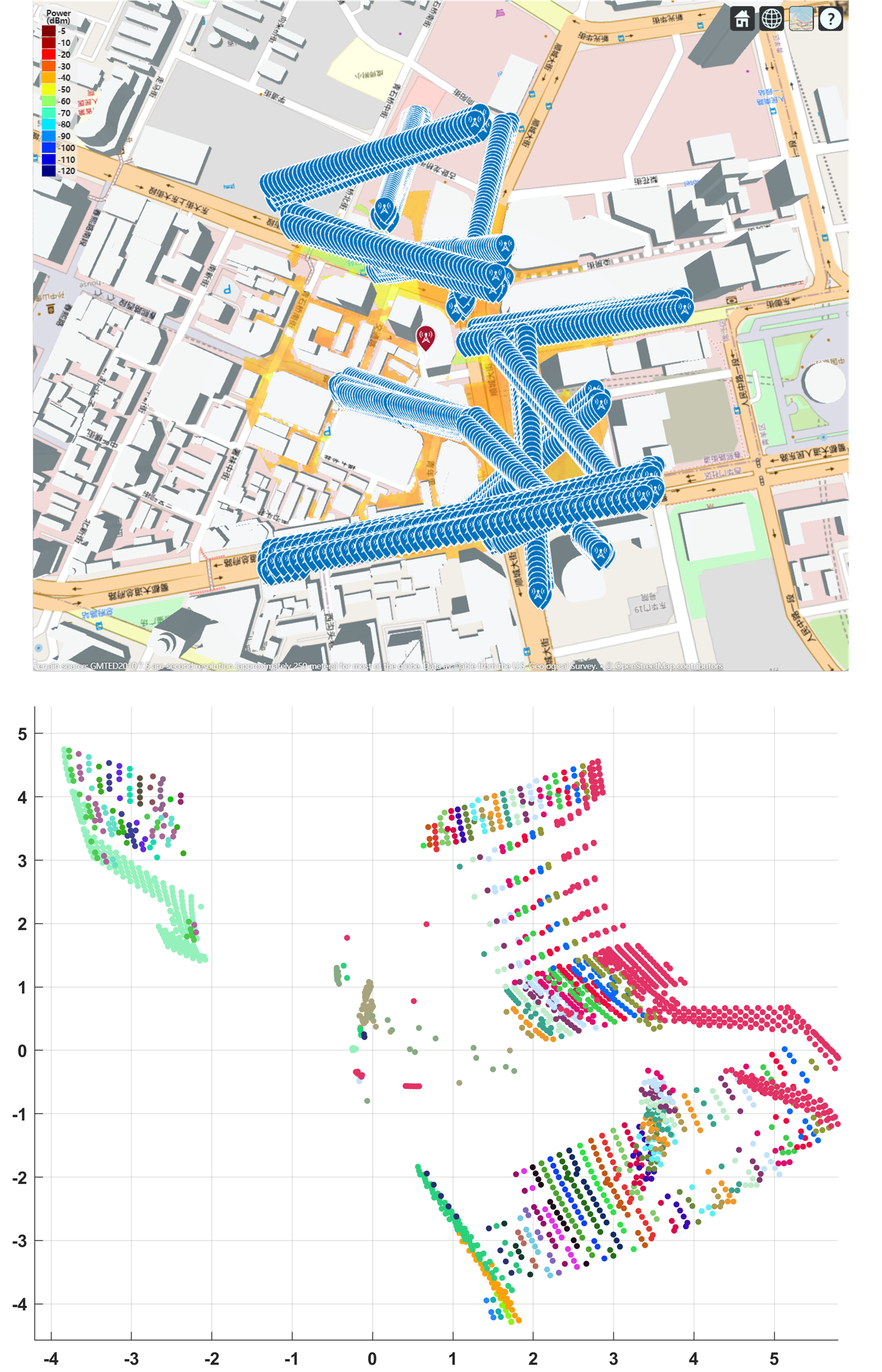}
    \caption{Data collection environmen and the channel chart constructed by our tripartite network.}
    \label{CC_results}
\end{figure}

\begin{figure}[tp] 
    \centering
    \includegraphics[width=9cm]{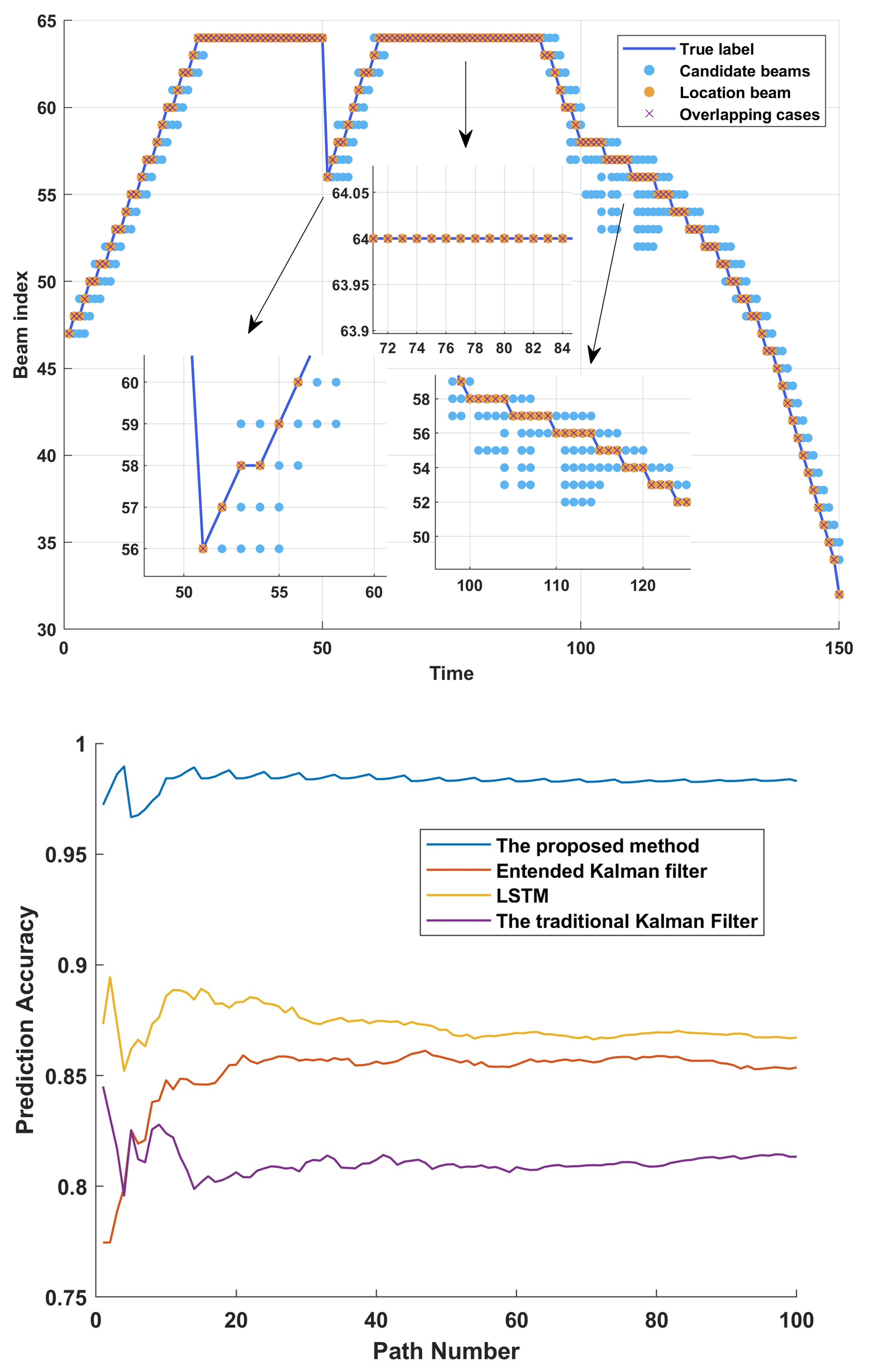}
    \caption{Running curves of real-time beam tracking algorithms and The accuracy comparison between the proposed method and other algorithms for 100 random paths under the same environment.}
    \label{runtime-performance-1}
\end{figure}

Fig. \ref{CC_results} shows the channel charts generated by our proposed algorithm (the second subfigure); the axes represent latent features extracted from CSI after dimensionality reduction. We select a two-dimensional channel chart ($D'=2$) to enable intuitive visualization while preserving spatial proximity. It also displays the terminal positions in both scenarios (the first subfigure); these data come from the simulation platform and show the actual user positions in the environment. It can be observed that, without relying on GPS information, the channel chart exhibits a certain degree of similarity to the terminal distribution. Angles are used to label the coordinates in the chart to distinguish between different types of coordinates.

Our proposed beam tracking algorithm displays information such as the located beam, actual beam, and dynamic candidate beam sets during operation. Fig.\ref{runtime-performance-1} illustrates the beam tracking along a path, showing how the predicted beams change over time.

From Fig.\ref{runtime-performance-1}, it is evident that the beam tracking algorithm proposed in this paper can dynamically adjust the candidate beam set. Between moments 1 and 26, the beam index increases persistently, covering the actual beam at each moment within the candidate beam set. From moments 27 to 51, the beam index remains stable at index 64, and during this period, the candidate beam set only includes the beam at index 64, maintaining high-quality communication at minimal cost. At moment 52, the scenario changes, causing a beam jump; however, the candidate beam set at moment 51 had already captured this situation by preemptively including the transitioning beam. After moment 92, the beam index continues to exhibit persistent changes. Note that between moments 100 and 114, the number of beams in the set increases due to the increased jitter and range of beam changes occurring after moment 114. Thus, the increase in the number of candidate beams during the interval from 100 to 114 is intended to prevent beam misalignment.





\begin{figure}[tp] 
    \centering
    \includegraphics[width=9.2cm]{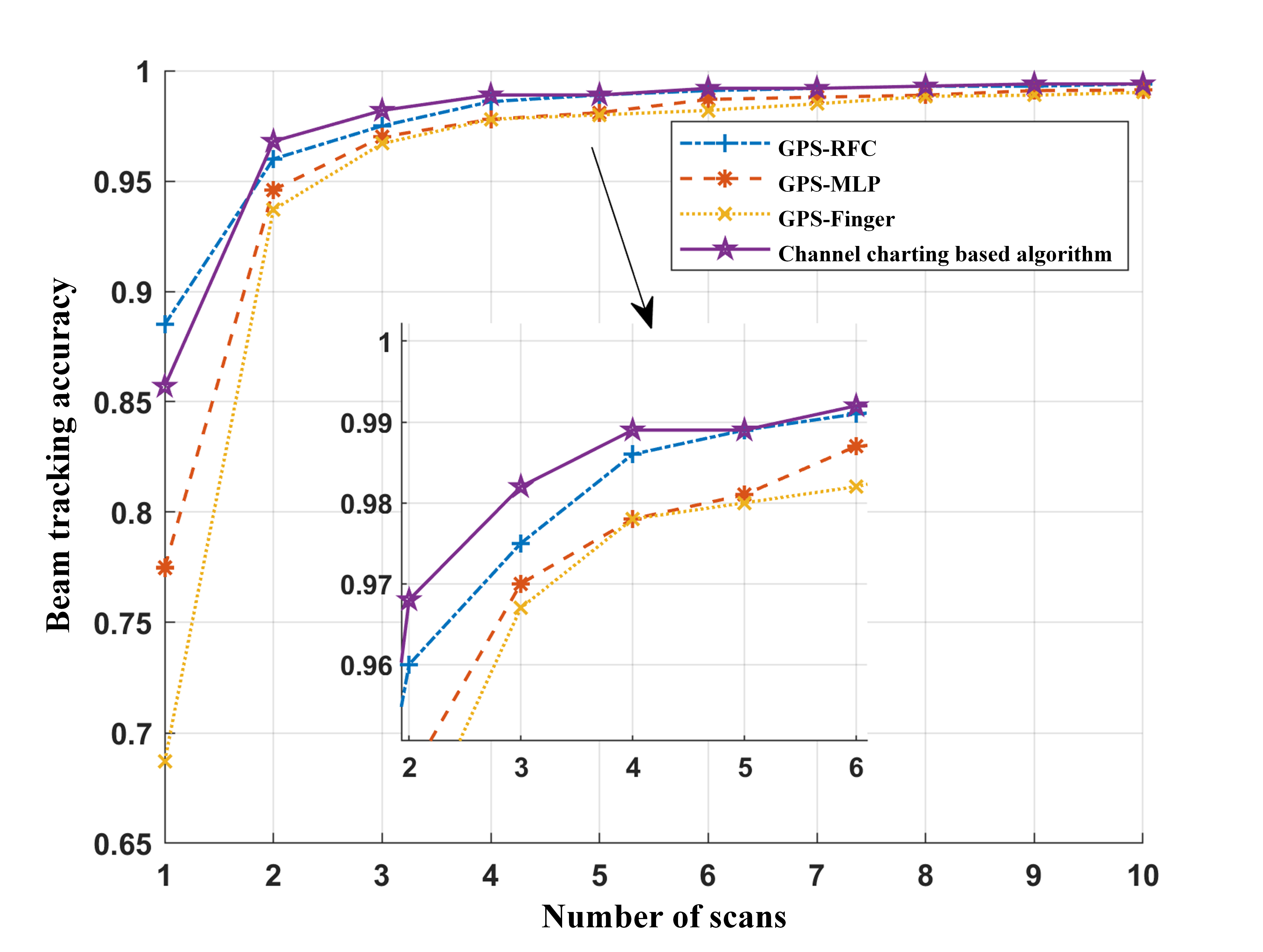}
    \caption{Comparison of the number of scans between the proposed algorithm and algorithms using location information. GPS-MLP/GPS-RFC/GPS-Finger use UE GPS/location as input (MLP classifier, Random Forest classifier, and location-fingerprint/nearest-neighbor method, respectively).}
    \label{fig:comparasion of diff algo}
\end{figure}

A comparison of the accuracy of different schemes \cite{liu_ekf-based_2019}\cite{zhang_tracking_2016}\cite{lim_deep_2021} is presented in Fig. \ref{fig:comparasion of diff algo}. We evaluated the prediction results of 200 paths using various methods. Traditional algorithms struggle to accurately predict the beam direction when switching between LOS and NLOS scenarios, despite the strong correlation between angle sequences and time sequences, as well as their inherent state transition relationships. In contrast, our method demonstrates high and stable accuracy while effectively capturing abrupt channel angle transitions. The proposed algorithm achieves over 98\% accuracy with an average of only 3 scans, while some algorithms require more than 6 scans. Statistical results show that our algorithm reduces the number of scans by up to 55.9\% compared to other algorithms at the same prediction accuracy level.

Compared to the Kalman-based method, our proposed algorithm avoids the cumulative errors. The performance of Kalman filters is highly sensitive to the accuracy of both the system model and the noise statistical model. When the system's mathematical model or the noise characteristics are misrepresented, the Kalman filter may fail to accurately reflect the underlying physical process, resulting in discrepancies between the observed data and the model, which can lead to cumulative errors over time. In contrast, the beam tracking algorithm we present is fundamentally data-driven, and thus does not suffer from these limitations.

\subsection{Timeliness analysis}
In the beam tracking algorithm proposed in this paper, the channel chart generation model requires periodic retraining to accommodate changes in the wireless channel environment and updates to the dataset. The retraining intervals for different CC methods vary significantly. Experimental results from \cite{howsoon-cc} indicate that when using PCA to generate the channel chart, the retraining interval is 30 seconds. We will conduct a simulation analysis on the timeliness of the beam tracking algorithm based on the channel chart presented in this paper.

\begin{figure*}[htp]
    \centering
    \includegraphics[width=18cm]{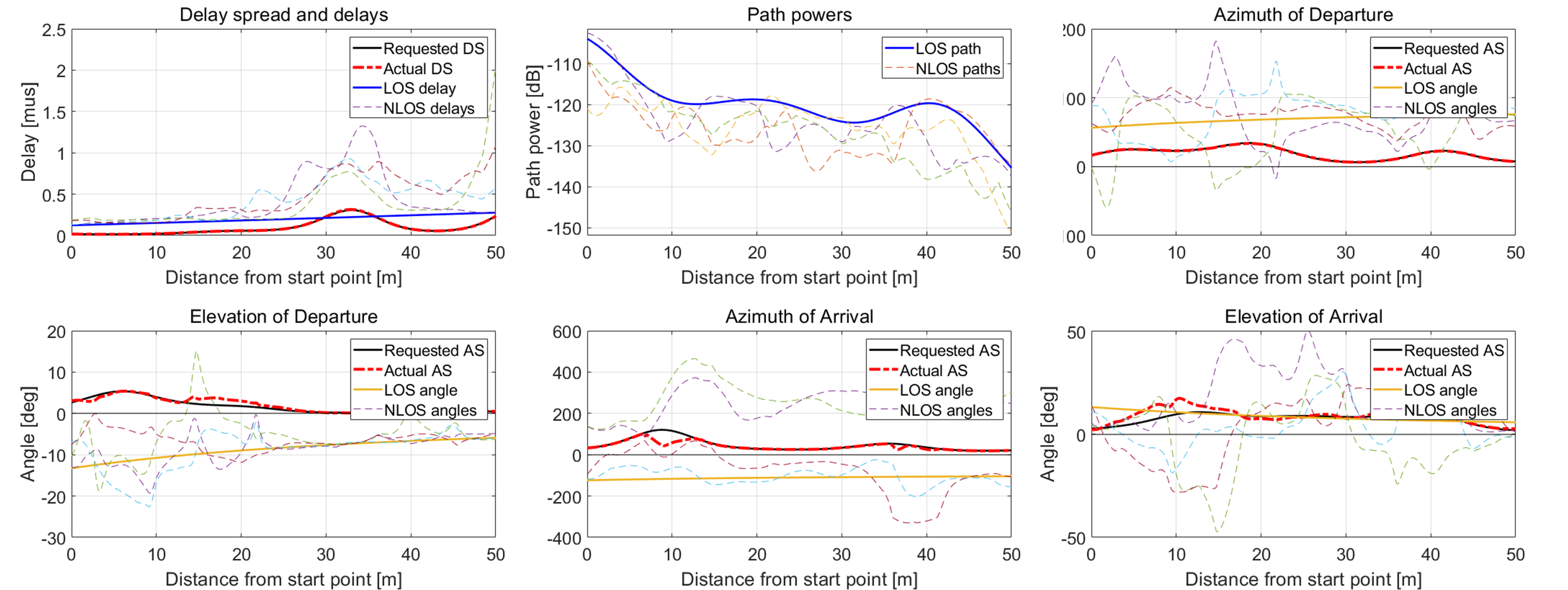}
    \caption{Simulation scene information generated by QuaDRiGa}
    \label{qua-scene-info}
\end{figure*}

To accommodate large-scale temporal variations, this study utilized the open-source channel model QuaDRiGa, setting a fixed transmitter and a randomly moving receiver, with scenarios that include both LOS and NLOS conditions. QuaDRiGa creates a time-varying channel by the movement of the terminal, segmenting a long path into different short intervals \cite{QuaDRiGa}. Within these intervals, the channel maintains generalized stationary properties with minimal angular changes. The length of each interval is determined by the decorrelation distance of large-scale fading. Empirical values are typically 20 meters for LOS environments and 45 meters for NLOS environments. Fig.\ref{qua-scene-info} presents the simulation environment information used to explore the timeliness of the algorithm, showing the CSI changes within the short intervals of the QuaDRiGa channel model. The environment is composed of several concatenated short intervals. The base station is set at a height of 10 meters, supports up to 5 maximum clusters, and has a small-scale decorrelation distance of 5 meters. 

\begin{figure}[tp]
    \centering
    \includegraphics[width=9.2cm]{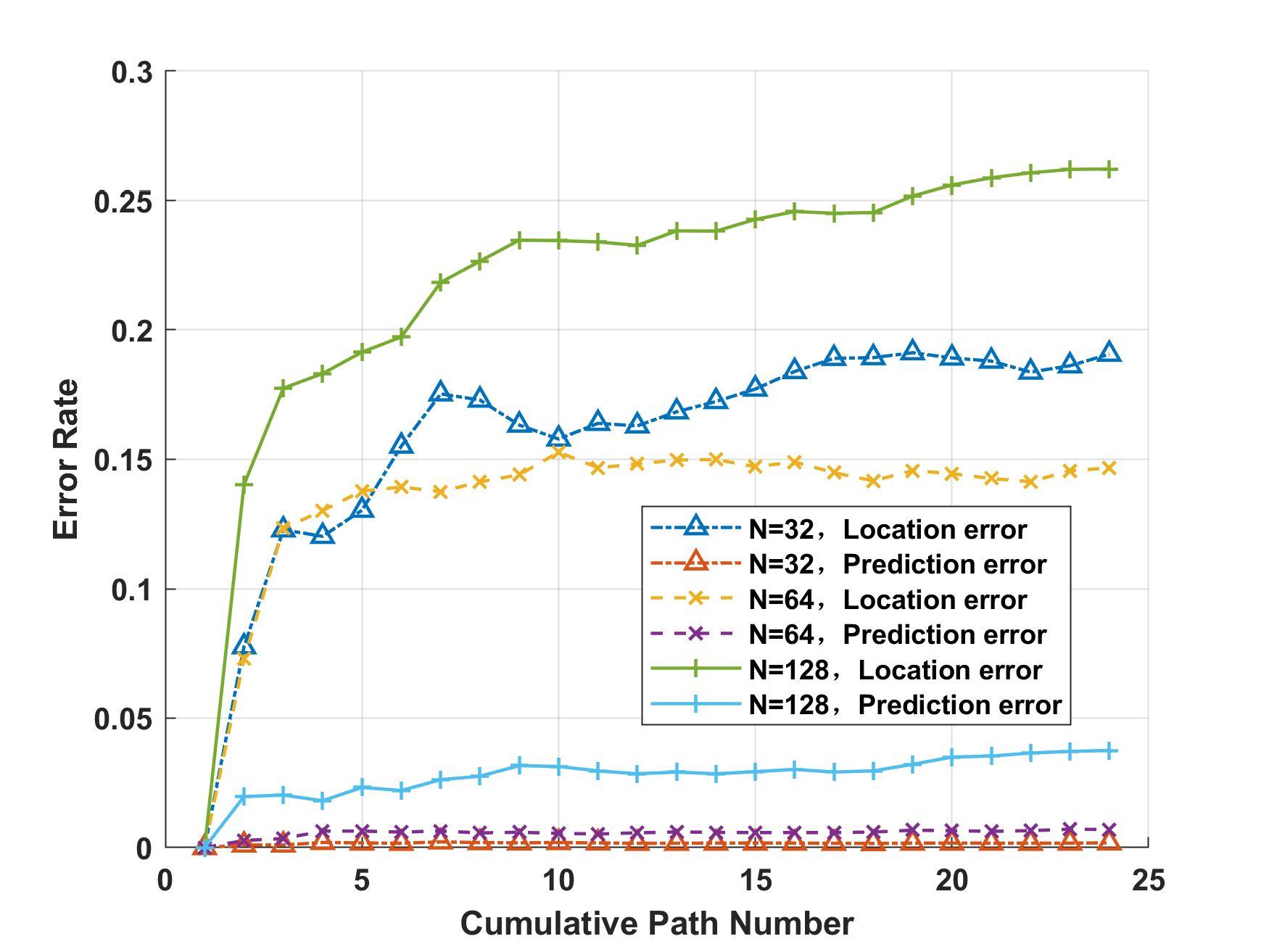}
    \caption{Channel The error rate of the beam tracking algorithm over the subsequent 24 path durations under the 3GPP standard secnario}
    \label{cc-valid-errorANDcc}
\end{figure}

We use the error rate to intuitively assess the timeliness of the channel chart, within a simulation scenario based on the 3GPP standard urban NLOS environment. At the initial moment, the channel chart is constructed, followed by continuous input of new data for prediction and error recording over the next 24 path durations. The error rates for the beam tracking algorithm under different codebook sizes are shown in Fig.\ref{cc-valid-errorANDcc}. The following results can be observed from the figure:
\begin{enumerate}
    \item As time progresses and the user terminal moves, both positioning and prediction errors gradually increase, with positioning errors becoming more pronounced. This indicates that the effectiveness of the channel chart diminishes over time, leading to increasing errors, although some usable information still remains.

    \item The simulation results did not show significant increases in prediction error, demonstrating the stability of the beam tracking scheme based on the channel chart proposed in this paper. Compared to \cite{howsoon-cc}, our test scenario covers a longer time span, with one path comprising 2004 terminal movement positions, with a 1-second interval between adjacent positions.

    \item In the simulation results, positioning errors were greater than prediction errors, with both types of errors showing a continuous upward trend. Taking the example of a codebook size of 128, positioning errors showed three significant increases after the 2nd, 12th, and 20th paths. However, this had little impact on prediction errors, indicating the good stability of our algorithm. After the 20th path, however, the values of both errors further increased.
\end{enumerate}

Based on the above observations, we believe that to maintain high prediction accuracy, the retraining interval for the channel chart-based beam tracking algorithm should be about $20 \times 2004 = 40,080$ seconds. In practical systems, it is necessary to consider the specific model and the amount of data involved.

\subsection{Different Configuration}
The beam tracking algorithm proposed in this paper is fundamentally a data-driven prediction problem. The configuration of the algorithm plays a significant role in the final prediction outcomes. Both the positioning and prediction errors are subject to fluctuations due to different algorithm configurations.

In the algorithm we propose, the number of dimensions in the input layer can be either 3 or 5, the origin $\mathbf{x}_{t}$ can equal $[\alpha_t^{rx},\alpha_t^{tx},\tau_t]$, $[\phi_t^{tx},\phi_t^{rx},\tau_t]$ or the combination of $\alpha$ and $\phi$, i.e., $\mathbf{x}_t = \left[\alpha_t^{rx},\phi_t^{tx},\alpha_t^{tx},\phi_t^{rx},\tau_t\right]$. The hidden layers decrease in size by a factor of 2 starting from the first layer, with subsequent layers reducing further by a factor of 3 to 5.

\begin{figure}[tp] 
    \centering
    \includegraphics[width=9cm]{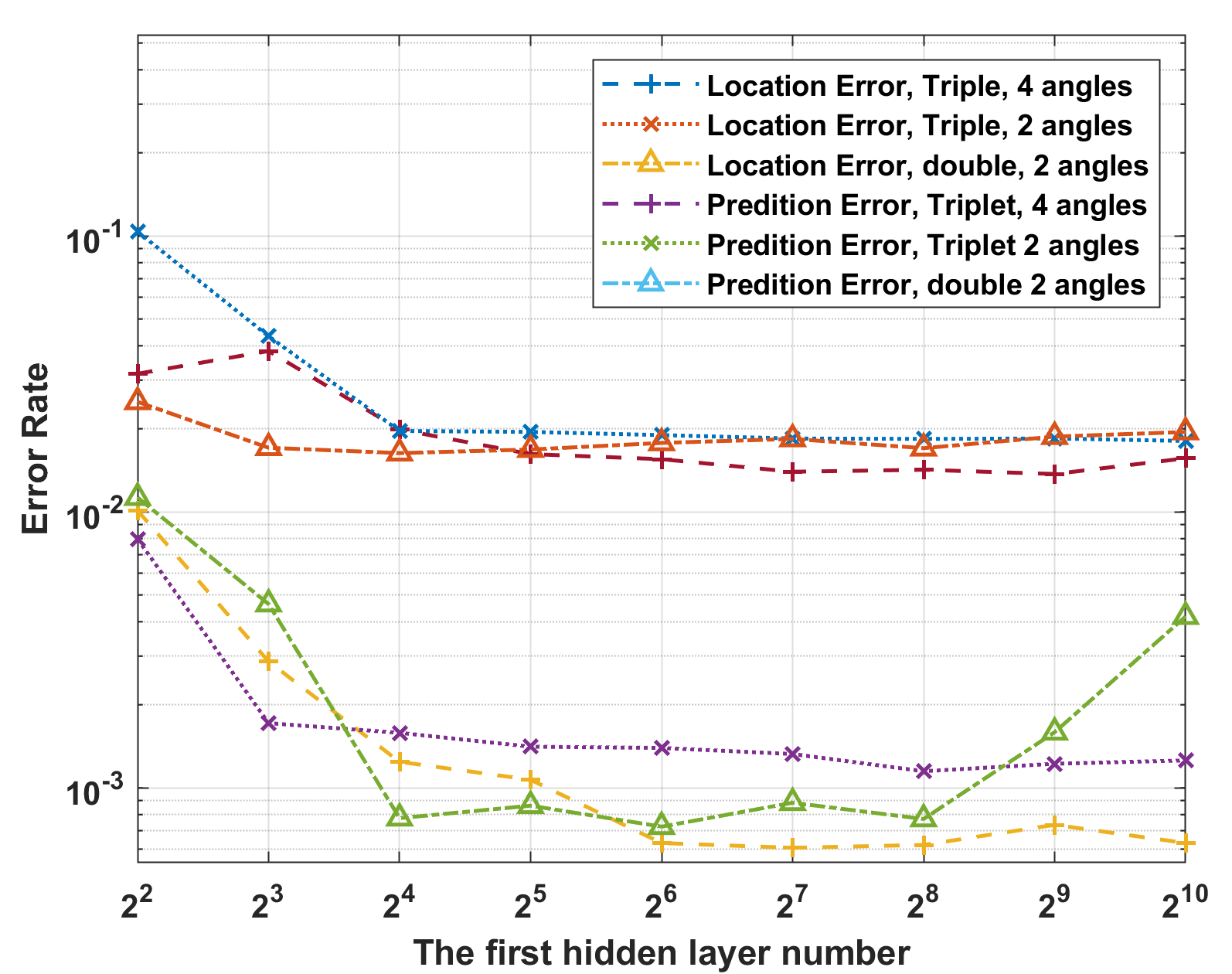}
    \caption{Beam tracking error rate under different input dimensions and hidden layers}
    \label{different-config-compare}
\end{figure}

The selection of hidden layers and input layers significantly impacts the final performance of the algorithm. Generally, networks with greater depth and more neurons demonstrate stronger fitting capabilities, which can be explained by the universal approximation theorem. However, the algorithm we propose is not entirely a curve fitting problem; therefore, it is necessary to compare the final errors to select the appropriate parameters. Fig.\ref{different-config-compare} displays the algorithmic errors under different input dimensions and numbers of hidden layers. From the curves, the following conclusions can be drawn.
\begin{enumerate}
    \item As the number of units in the first hidden layer increases, the positioning error tends to decrease. With an input of four angles and a hidden layer size of 256, the positioning error is minimized. Further increasing the dimension of the hidden layers leads to a rebound in error.

    \item When the input information includes four angles, the prediction error is smaller compared to other scenarios. The minimum positioning error is achieved when the number of the hidden layer number is set to 256.

    \item Within the range of 4 to 32 units in the first hidden layer, the decrease in both positioning and prediction errors is particularly pronounced.
    
\end{enumerate}

\subsection{Field tests}

To validate the feasibility of the beam tracking algorithm, we constructed a 26 GHz real millimeter-wave communication test system and designed the corresponding software systems to enable the algorithm to run on hardware. Both the transmitter and receiver are equipped with phased array antennas consisting of 4 rows, each with 16 elements. The beam direction is configured via digital I/O to the antenna's phase shifters.

\begin{figure}[tp] 
    \centering
    \includegraphics[width=8.8cm]{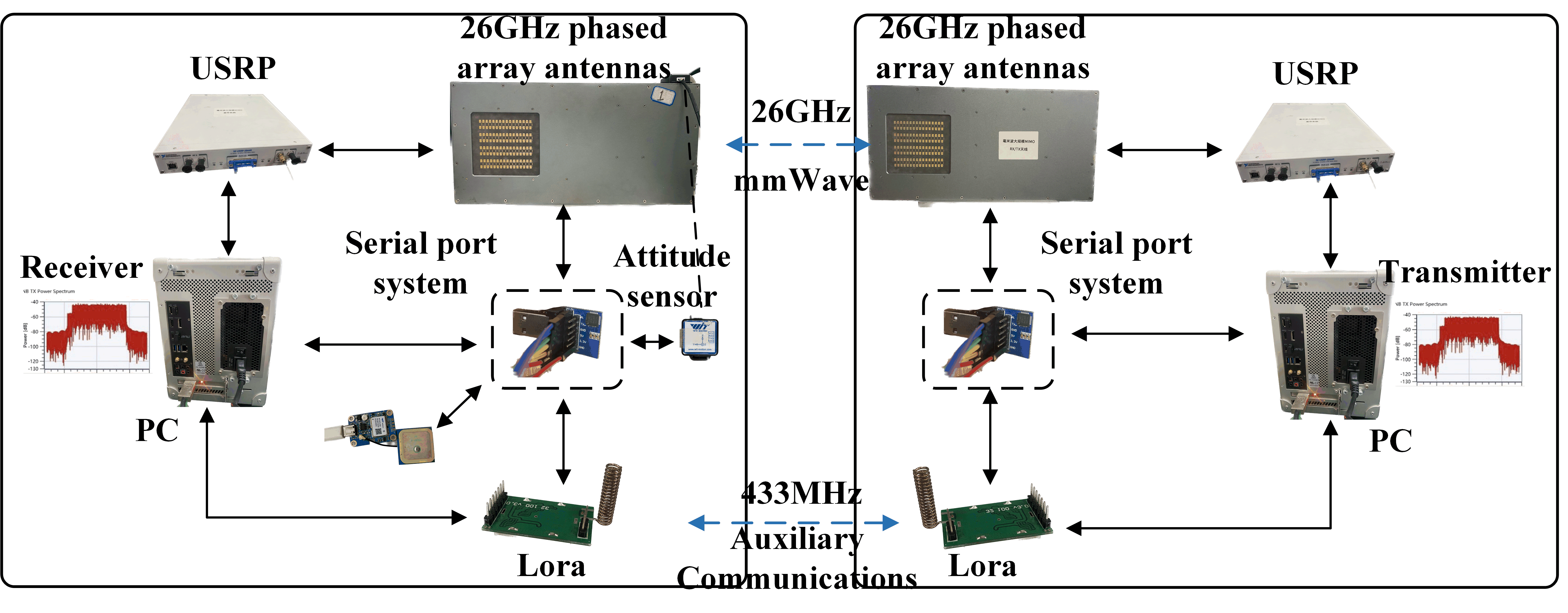}
    \includegraphics[width=8.8cm]{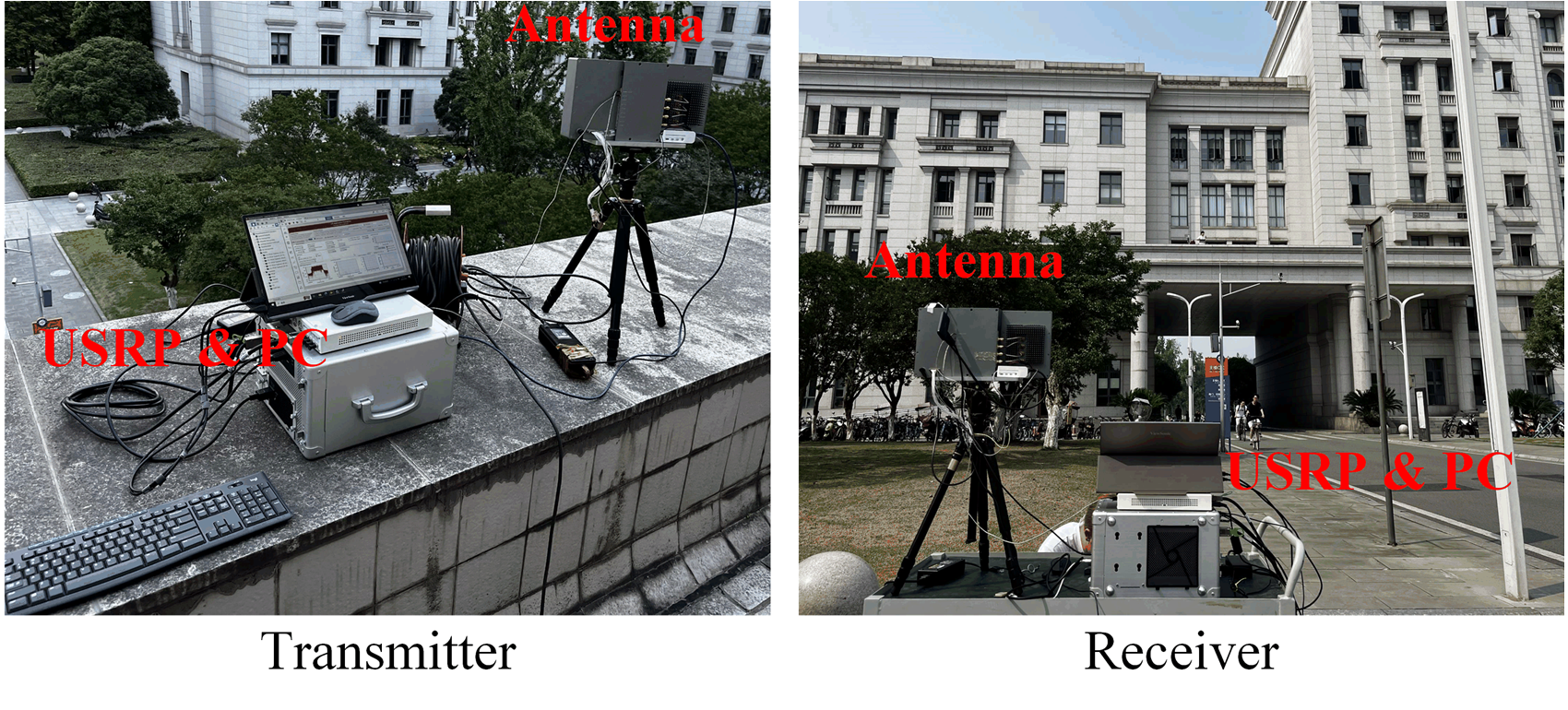}
    \caption{Communication system equipment architecture of the field test, and our test scenario}
    \label{26GHz-system}
\end{figure}

Fig.\ref{26GHz-system} shows the hardware framework of the test system we constructed. The beam tracking algorithm operates on a host computer controlled by a central processor. The host computer connects directly to the software-defined radio that handles the transmit and receive intermediate frequency signals, using the energy values of these signals to determine the final predicted beam direction. In our setup, the transmitter is fixed, while the receiver can move or rotate. For this reason, we use a posture sensor to record the rotation angles, enabling the transformed coordinate system to be realigned with the original coordinate system after rotation. Table \ref{tab:table5} summarizes the main hardware specification parameters. Our test scenario is shown in Fig.\ref{26GHz-system}.



\begin{table}[!t]
\caption{Hardware Parameters of Algorithm Prototype \label{tab:table5}}
\centering
\begin{tabular}{|c|c|c|c|}
\hline
Antenna Parameters & Value & USRP Parameters & Value \\
\hline
Baud Rate & 115200 & Frequency Range & 30 M-6 GHz \\
\hline
Channel Size & $4\times16$ & \makecell[c]{Physical Resource\\Block}  & TS 36.211\\
\hline
\makecell[c]{Array Element \\ Spacing}  & 5.4 mm & Modulation & \makecell[c]{QPSK or \\ 16/64 QAM}  \\
\hline
Clock Frequency & 100 MHz & Sample Rate & 192 Mps\\
\hline
Radiant Power & 45 dBm & FFT Points & 2048\\
\hline
Data Protocol & UART & \makecell[c]{Subcarrier spacing \\ and bandwidth} & \makecell[c]{15 kHz \\ 20 MHz}\\
\hline
\end{tabular}

\end{table}

\begin{table}[!t]
\caption{Summary of Training Data, Model Architecture, and Optimization}
\centering
\begin{tabular}{|p{2cm}|p{6cm}|}
\hline
\textbf{Category} & \textbf{Content} \\
\hline
Training data &
Source: Ray-tracing platform (Wireless InSite-like) \newline
Split: Train/val/test split is performed temporally/spatially.
\\
\hline
Model architecture &
Encoder-decoder (autoencoder-style). \newline
Network depth: 6 layers; decoder is a symmetric mirror of the encoder. \newline
Input dimensionality: 5 (as above); latent/channel-chart dimensionality: 2. \newline
Loss: Triplet loss + decoding loss. As described in (\ref{cost2}).
\\
\hline
Optimization &
Optimizer: Adam. \newline
Learning rate: 1e-4 \newline
Batch size: 256; number of epochs: 200
\\
\hline
\end{tabular}

\end{table}

\begin{figure}[!t]
    \centering
    \includegraphics[width=9cm]{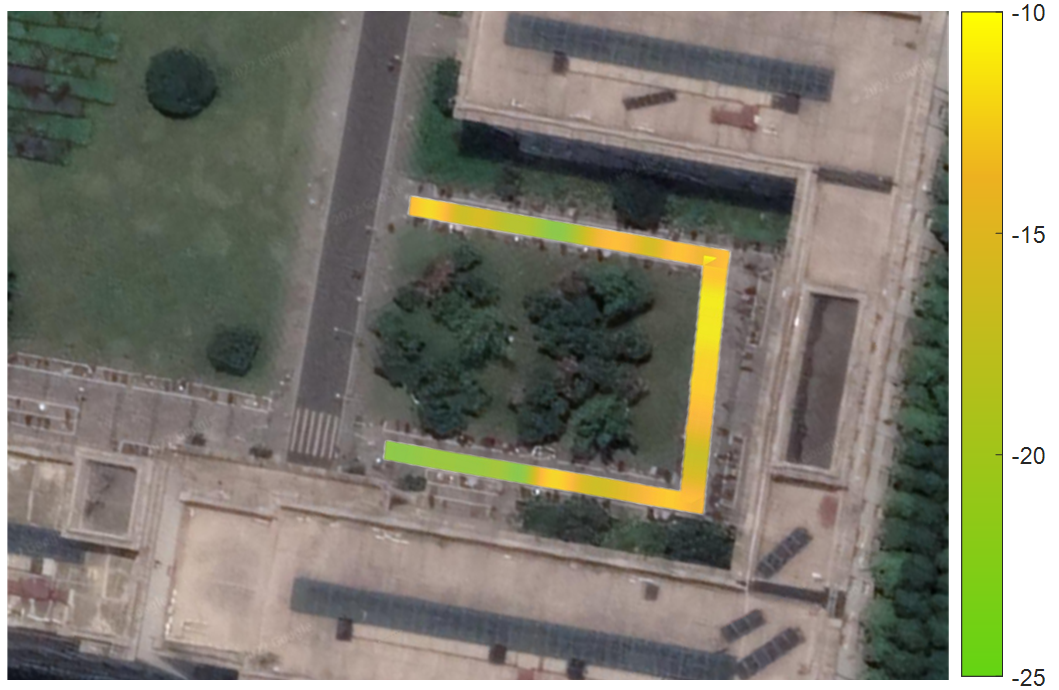}
    \caption{Signal power (dB) obtained from the field tests}
    \label{paths_energy}
\end{figure}

As shown in Fig.\ref{paths_energy}, we simulated a mobile terminal. To test the data sets back in the simulated environment , we selected a path with several NLOS paths to perform our proposed algorithm. Statistically speaking, the received signal power ranged from -23.46 to -10.78 dB. Compared with the performance in beam alignment, the average value of the beam tracking process is -13.82 dB, which is relatively high.

In our experimental setup, we selected a mixed LOS and NLOS scenario. In the case of signal
attenuation in free space, with beam alignment, the maximum received energy after environmental attenuation was measured at -10.78 dB. During receiver movement, NLOS conditions are frequent, resulting in further signal degradation. We consider -13.82 dB to be an acceptable threshold for received energy.

\section{Conclusion}
This study introduces an effective and consistent beam-tracking algorithm that relies on codebooks and leverages channel charting for dynamic conditions. The data-driven deep neural network is trained using channel state information CSI to produce low-dimensional channel charts. From these charts, the algorithm identifies multiple candidate beams for each tracking instance and selects the optimal beam through scanning, significantly decreasing the number of searches needed. Verification through simulations and real-world testing reveals that our method lessens search complexity while sustaining robustness in mobile mmWave communication scenarios.

{\small
\bibliographystyle{IEEEtran}
\bibliography{bib/beamtracking}

\begin{thebibliography}{10}
\providecommand{\url}[1]{#1}
\csname url@samestyle\endcsname
\providecommand{\newblock}{\relax}
\providecommand{\bibinfo}[2]{#2}
\providecommand{\BIBentrySTDinterwordspacing}{\spaceskip=0pt\relax}
\providecommand{\BIBentryALTinterwordstretchfactor}{4}
\providecommand{\BIBentryALTinterwordspacing}{\spaceskip=\fontdimen2\font plus
\BIBentryALTinterwordstretchfactor\fontdimen3\font minus
  \fontdimen4\font\relax}
\providecommand{\BIBforeignlanguage}[2]{{%
\expandafter\ifx\csname l@#1\endcsname\relax
\typeout{** WARNING: IEEEtran.bst: No hyphenation pattern has been}%
\typeout{** loaded for the language `#1'. Using the pattern for}%
\typeout{** the default language instead.}%
\else
\language=\csname l@#1\endcsname
\fi
#2}}
\providecommand{\BIBdecl}{\relax}
\BIBdecl

\bibitem{fast_beam_alignment}
X.~Li, J.~Fang, H.~Duan, Z.~Chen, and H.~Li, ``Fast beam alignment for
  millimeter wave communications: A sparse encoding and phaseless decoding
  approach,'' \emph{IEEE Transactions on Signal Processing}, vol.~67, no.~17,
  pp. 4402--4417, 2019.

\bibitem{noncooperative}
T.~L. Marzetta, ``Noncooperative cellular wireless with unlimited numbers of
  base station antennas,'' \emph{IEEE Transactions on Wireless Communications},
  vol.~9, no.~11, pp. 3590--3600, 2010.

\bibitem{8458146}
M.~Giordani, M.~Polese, A.~Roy, D.~Castor, and M.~Zorzi, ``A tutorial on beam
  management for 3gpp nr at mmwave frequencies,'' \emph{IEEE Communications
  Surveys and Tutorials}, vol.~21, no.~1, pp. 173--196, 2019.

\bibitem{8757185}
Y.~Guo, Z.~Wang, M.~Li, and Q.~Liu, ``Machine learning based mmwave channel
  tracking in vehicular scenario,'' in \emph{2019 IEEE International Conference
  on Communications Workshops (ICC Workshops)}, 2019, pp. 1--6.

\bibitem{7063467}
J.~He, T.~Kim, H.~Ghauch, K.~Liu, and G.~Wang, ``Millimeter wave mimo channel
  tracking systems,'' in \emph{2014 IEEE Globecom Workshops (GC Wkshps)}, 2014,
  pp. 416--421.

\bibitem{7510902}
C.~Zhang, D.~Guo, and P.~Fan, ``Tracking angles of departure and arrival in a
  mobile millimeter wave channel,'' in \emph{2016 IEEE International Conference
  on Communications (ICC)}, 2016, pp. 1--6.

\bibitem{zhang2019codebook}
D.~Zhang, A.~Li, M.~Shirvanimoghaddam, P.~Cheng, Y.~Li, and B.~Vucetic,
  ``Codebook-based training beam sequence design for millimeter-wave tracking
  systems,'' \emph{IEEE Transactions on Wireless Communications}, vol.~18,
  no.~11, pp. 5333--5349, 2019.

\bibitem{zhang2021training}
D.~Zhang, A.~Li, C.~Pradhan, J.~Li, B.~Vucetic, and Y.~Li, ``Training beam
  sequence design for multiuser millimeter wave tracking systems,'' \emph{IEEE
  Transactions on Communications}, vol.~69, no.~10, pp. 6939--6955, 2021.

\bibitem{ning2021unified}
B.~Ning, Z.~Chen, Z.~Tian, C.~Han, and S.~Li, ``A unified 3d beam training and
  tracking procedure for terahertz communication,'' \emph{IEEE Transactions on
  Wireless Communications}, vol.~21, no.~4, pp. 2445--2461, 2021.

\bibitem{liang2023millimetre}
C.~Liang, J.~Kuang, F.~Wu, and J.~Chen, ``Millimetre-wave beam tracking: An
  intelligent machine learning and kalman filter fusion technology,''
  \emph{IEEE Transactions on Cognitive Communications and Networking}, 2023.

\bibitem{burghal2019machine}
D.~Burghal, N.~A. Abbasi, and A.~F. Molisch, ``A machine learning solution for
  beam tracking in mmwave systems,'' in \emph{2019 53rd Asilomar Conference on
  Signals, Systems, and Computers}.\hskip 1em plus 0.5em minus 0.4em\relax
  IEEE, 2019, pp. 173--177.

\bibitem{dehkordi2021adaptive}
S.~K. Dehkordi, M.~Kobayashi, and G.~Caire, ``Adaptive beam tracking based on
  recurrent neural networks for mmwave channels,'' in \emph{2021 IEEE 22nd
  International Workshop on Signal Processing Advances in Wireless
  Communications (SPAWC)}.\hskip 1em plus 0.5em minus 0.4em\relax IEEE, 2021,
  pp. 1--5.

\bibitem{o49}
R.~Di~Taranto, S.~Muppirisetty, R.~Raulefs, D.~Slock, T.~Svensson, and
  H.~Wymeersch, ``Location-aware communications for 5g networks: How location
  information can improve scalability, latency, and robustness of 5g,''
  \emph{IEEE Signal Processing Magazine}, vol.~31, no.~6, pp. 102--112, 2014.

\bibitem{o50}
R.~Deng, S.~Chen, S.~Zhou, Z.~Niu, and W.~Zhang, ``Channel fingerprint based
  beam tracking for millimeter wave communications,'' \emph{IEEE Communications
  Letters}, vol.~24, no.~3, pp. 639--643, 2019.

\bibitem{ferrand_triplet-based_2021}
\BIBentryALTinterwordspacing
P.~Ferrand, A.~Decurninge, L.~G. Ordonez, and M.~Guillaud, ``Triplet-{Based}
  {Wireless} {Channel} {Charting}: {Architecture} and {Experiments},''
  \emph{IEEE Journal on Selected Areas in Communications}, vol.~39, no.~8, pp.
  2361--2373, Aug. 2021. [Online]. Available:
  \url{https://ieeexplore.ieee.org/document/9448128/}
\BIBentrySTDinterwordspacing

\bibitem{9306087}
P.~Kazemi, H.~Al-Tous, C.~Studer, and O.~Tirkkonen, ``Snr prediction in
  cellular systems based on channel charting,'' in \emph{2020 IEEE Eighth
  International Conference on Communications and Networking (ComNet)}, 2020,
  pp. 1--8.

\bibitem{kazemi_channel_2022}
\BIBentryALTinterwordspacing
------, ``\BIBforeignlanguage{en-US}{Channel {Charting} {Assisted} {Beam}
  {Tracking}},'' in \emph{\BIBforeignlanguage{en-US}{2022 {IEEE} 95th
  {Vehicular} {Technology} {Conference}: ({VTC2022}-{Spring})}}.\hskip 1em plus
  0.5em minus 0.4em\relax Helsinki, Finland: IEEE, Jun. 2022, pp. 1--5.
  [Online]. Available: \url{https://ieeexplore.ieee.org/document/9860709/}
\BIBentrySTDinterwordspacing

\bibitem{ponnada2021best}
T.~Ponnada, P.~Kazemi, H.~Al-Tous, Y.-C. Liang, and O.~Tirkkonen, ``Best beam
  prediction in non-standalone mm wave systems,'' in \emph{2021 Joint European
  Conference on Networks and Communications \& 6G Summit (EuCNC/6G
  Summit)}.\hskip 1em plus 0.5em minus 0.4em\relax IEEE, 2021, pp. 532--537.

\bibitem{our-pre-work}
J.~Zhang, F.~Wu, Y.~Yang, and J.~Chen, ``Beam tracking: A channel charting and
  neighborhood search based method,'' in \emph{GLOBECOM 2023-2023 IEEE Global
  Communications Conference}.\hskip 1em plus 0.5em minus 0.4em\relax IEEE,
  2023, pp. 807--812.

\bibitem{lim2021deep}
S.~H. Lim, S.~Kim, B.~Shim, and J.~W. Choi, ``Deep learning-based beam tracking
  for millimeter-wave communications under mobility,'' \emph{IEEE Transactions
  on Communications}, vol.~69, no.~11, pp. 7458--7469, 2021.

\bibitem{brady2013beamspace}
J.~Brady, N.~Behdad, and A.~M. Sayeed, ``Beamspace mimo for millimeter-wave
  communications: System architecture, modeling, analysis, and measurements,''
  \emph{IEEE Transactions on Antennas and Propagation}, vol.~61, no.~7, pp.
  3814--3827, 2013.

\bibitem{karacora2023event}
Y.~Karacora, C.~Chaccour, A.~Sezgin, and W.~Saad, ``Event-based beam tracking
  with dynamic beamwidth adaptation in terahertz (thz) communications,''
  \emph{IEEE Transactions on Communications}, vol.~71, no.~10, pp. 6195--6210,
  2023.

\bibitem{zhao2024lstm}
Y.~Zhao, X.~Zhang, X.~Gao, K.~Yang, Z.~Xiong, and Z.~Han, ``Lstm-based
  predictive mmwave beam tracking via sub-6 ghz channels for v2i
  communications,'' \emph{IEEE Transactions on Communications}, 2024.

\bibitem{studer-vtc}
P.~Kazemi, H.~Al-Tous, C.~Studer, and O.~Tirkkonen, ``Channel charting assisted
  beam tracking,'' in \emph{2022 IEEE 95th Vehicular Technology
  Conference:(VTC2022-Spring)}.\hskip 1em plus 0.5em minus 0.4em\relax IEEE,
  2022, pp. 1--5.

\bibitem{hornik1989multilayer}
K.~Hornik, M.~Stinchcombe, and H.~White, ``Multilayer feedforward networks are
  universal approximators,'' \emph{Neural networks}, vol.~2, no.~5, pp.
  359--366, 1989.

\bibitem{auto-encoder}
P.~Ghosh, M.~S. Sajjadi, A.~Vergari, M.~Black, and B.~Sch{\"o}lkopf, ``From
  variational to deterministic autoencoders,'' \emph{arXiv preprint
  arXiv:1903.12436}, 2019.

\bibitem{studer_channel_2018}
\BIBentryALTinterwordspacing
C.~Studer, S.~Medjkouh, E.~Gonultas, T.~Goldstein, and O.~Tirkkonen, ``Channel
  {Charting}: {Locating} {Users} {Within} the {Radio} {Environment} {Using}
  {Channel} {State} {Information},'' \emph{IEEE Access}, vol.~6, pp.
  47\,682--47\,698, 2018. [Online]. Available:
  \url{https://ieeexplore.ieee.org/document/8444621/}
\BIBentrySTDinterwordspacing

\bibitem{schroff_facenet_2015}
\BIBentryALTinterwordspacing
F.~Schroff, D.~Kalenichenko, and J.~Philbin, ``{FaceNet}: {A} unified embedding
  for face recognition and clustering,'' in \emph{2015 {IEEE} {Conference} on
  {Computer} {Vision} and {Pattern} {Recognition} ({CVPR})}.\hskip 1em plus
  0.5em minus 0.4em\relax Boston, MA, USA: IEEE, Jun. 2015, pp. 815--823.
  [Online]. Available: \url{http://ieeexplore.ieee.org/document/7298682/}
\BIBentrySTDinterwordspacing

\bibitem{introduction_to_algo}
T.~H. Cormen, C.~E. Leiserson, R.~L. Rivest, and C.~Stein, \emph{Introduction
  to algorithms}.\hskip 1em plus 0.5em minus 0.4em\relax MIT press, 2022.

\bibitem{liu_ekf-based_2019}
F.~Liu, P.~Zhao, and Z.~Wang, ``{EKF}-{Based} {Beam} {Tracking} for {mmWave}
  {MIMO} {Systems},'' \emph{IEEE Communications Letters}, vol.~23, no.~12, pp.
  2390--2393, Dec. 2019.

\bibitem{zhang_tracking_2016}
C.~Zhang, D.~Guo, and P.~Fan, ``Tracking angles of departure and arrival in a
  mobile millimeter wave channel,'' in \emph{2016 {IEEE} {International}
  {Conference} on {Communications} ({ICC})}.\hskip 1em plus 0.5em minus
  0.4em\relax Kuala Lumpur, Malaysia: IEEE, May 2016, pp. 1--6.

\bibitem{lim_deep_2021}
S.~H. Lim, S.~Kim, B.~Shim, and J.~W. Choi, ``Deep {Learning}-{Based} {Beam}
  {Tracking} for {Millimeter}-{Wave} {Communications} {Under} {Mobility},''
  \emph{IEEE Transactions on Communications}, vol.~69, no.~11, pp. 7458--7469,
  Nov. 2021.

\bibitem{howsoon-cc}
Z.~Wang, Y.~Xu, M.~Zhao, S.~Zhang, and G.~He, ``How soon will channel charting
  be inapplicable in user moving scenarios?: Some answers,'' in \emph{GLOBECOM
  2022-2022 IEEE Global Communications Conference}.\hskip 1em plus 0.5em minus
  0.4em\relax IEEE, 2022, pp. 1350--1355.

\bibitem{QuaDRiGa}
S.~Jaeckel, L.~Raschkowski, K.~B{\"o}rner, and L.~Thiele, ``Quadriga: A 3-d
  multi-cell channel model with time evolution for enabling virtual field
  trials,'' \emph{IEEE transactions on antennas and propagation}, vol.~62,
  no.~6, pp. 3242--3256, 2014.

\end{thebibliography}
}

\end{document}